\documentclass[times,final]{elsarticle}
\usepackage{doi}
\usepackage{subfig}    
\usepackage{graphicx}  
\usepackage{amsmath} 
\usepackage{jcomp}
\usepackage{framed,multirow}
\usepackage{tabularx}
\usepackage{setspace}
\usepackage{amssymb}
\usepackage{amsmath,amssymb}
\usepackage{latexsym}
\usepackage{lineno}

\begin{document}

\noindent{\LARGE{\textbf{Characteristic boundary conditions for magnetohydrodynamic equations}}}%

\noindent{\large{P. Makaremi-Esfarjani, A. Najafi-Yazdi}}\\
\noindent{\footnotesize Mechanical Engineering Department, McGill University, Montr\'eal, QC, Canada}

\section*{\large{Abstract}}
In the present study, a characteristic-based boundary condition scheme is developed for the compressible magnetohydrodynamic (MHD) equations in the general curvilinear coordinate system, which is an extension of the characteristic boundary scheme for the Navier-Stokes equations. The eigenstructure and the complete set of characteristic waves are derived for the ideal MHD equations in
general curvilinear coordinates $(\xi, \eta, \zeta)$. The characteristic boundary conditions are derived and implemented in a high-order MHD solver where the sixth-order compact scheme is used for the spatial discretization. The fifth-order Weighted Essentially Non-Oscillatory (WENO) scheme is also employed for the spatial discretization of problems with discontinuities. In our MHD solver, the fourth-order Runge-Kutta scheme is utilized for time integration. The characteristic boundary scheme is first verified for the non-magnetic (i.e., $\mathbf{B}=\textbf{0}$) Sod shock tube problem. Then, various in-house test cases are designed to examine the derived MHD characteristic boundary scheme for three different types of boundaries: non-reflecting inlet and outlet, solid wall, and single characteristic wave injection. The numerical examples demonstrate the accuracy and robustness of the MHD characteristic boundary scheme.\\
{\textbf{Keywords:} \textit{Magnetohydrodynamics, MHD, Characteristic boundary, Boundary conditions, Characteristic waves, Non-reflecting inlet/outlet boundary}}
 
\section{Introduction}
Imposing appropriate boundary conditions is of crucial importance to the accuracy and convergence of numerical results. Characteristic boundary conditions are among the most accurate boundary conditions used in typical Computational Fluid Dynamic (CFD) problems involving transient compressible flows. Such boundary conditions are constructed based on identifying the outgoing and incoming waves from and to the computational domain. The amplitude of the incoming waves are set to appropriate values to impose the desired boundary condition. In this scheme, no extrapolation is used since different boundary conditions can be applied by using characteristic relations of the crossing waves at the boundary.
\par
Characteristic boundary conditions for the Euler equations were extensively studied by Kreiss \citep{Kreiss1970}, Engquist and Majda \citep{Engquist&Majda1977}, and Hirsch \citep{Hirsch1988}. This scheme was also studied for the Navier-Stokes equations by Poinsot and Lele \citep{Poinsot&Lele1992}. Poinsot and Lele showed that the characteristic method has higher precision and stability in comparison with other classical approaches such as extrapolation and simplified Riemann invariants \citep{Poinsot&Lele1992,Cimino&Krause2016}. Moreover, Thompson \citep{Thompson1987} studied the time-dependent characteristic boundary conditions for a general hyperbolic system.
\par
 Most numerical tools to solve MHD equations involve periodic computational domains, which considerably limit the application of such numerical simulations. One of the most practical boundary conditions is the non-reflecting inlet/outlet that can be readily implemented using the characteristic boundary scheme. Non-reflecting boundary conditions for compressible flows were first studied in the work of Bayliss and Turkel \citep{Bayliss&Turkel1982} by using the perturbation approach. In their study, a perturbation was expressed in terms of the characteristic waves in order to impose boundary conditions such that it would prevent the formation of incoming waves. The study of Bayliss and Turkel was followed by the work of Engquist and Majda \citep{Engquist&Majda1977,Engquist&Majda1979}, which developed the characteristic non-reflecting boundary scheme for linear acoustic and elastic waves. Later, Hedstrom \citep{Hedstrom1979} proposed the non-reflecting boundary scheme for general nonlinear hyperbolic systems as well. Hedstrom \citep{Hedstrom1979} suggested that a non-reflecting boundary condition involves setting the amplitude of the incoming waves to zero. Jiang et al. \citep{Jiang&Shan1999} further developed the non-reflecting boundary conditions in the general curvilinear coordinates for compressible flows. Here, we expand the idea of the non-reflecting boundary conditions to the ideal MHD equations.
\par
The application of characteristic boundary conditions for MHD problems has been somewhat limited. Cimino et al. \citep{Cimino&Krause2016} introduced the characteristic boundary scheme for the one-dimensional MHD equations in the cartesian coordinate system. They tried to benchmark two types of boundary conditions, open end and solid wall, for the Brio--Wu shock tube problem by comparing their results to that obtained from a zeroth-order extrapolated boundary condition scheme. This comparison showed some discrepancies that they suggested might be due to nonphysical boundary conditions, warranting further studies \citep{Cimino&Krause2016}.
\par
The goal of the present work is to introduce a set of characteristic boundary conditions for numerical simulation of the compressible MHD equations, especially on curvilinear grids. Here, the formal derivation of the characteristic boundary conditions for the numerical simulation of the MHD equations is examined in detail. The complete eigenstructure of the MHD equations is calculated and a set of benchmark problems is designed to verify the accuracy of the implemented boundary conditions.

\section{Governing ideal MHD equations}

The magnetohydrodynamic set of equations is a fluid model of plasma which develops a macroscopic description of its dynamics. In this model, plasma is assumed as a continuous medium where hydrodynamic length scales are much larger than microscopic ones, namely the particle mean free path and the gyro-radius. Similarly, hydrodynamic time scales are considered to be much greater than the mean free time and the gyro-period. Using these assumptions, fluid dynamic equations can be applied to the motion of plasma. These equations should be coupled with the Maxwell's equations to close the system of equations. The governing equations considered throughout this study are ideal MHD equations, in which all dissipative processes such as viscosity, electrical resistivity, and thermal conductivity are neglected.
\par
The ideal MHD equations can be written in the general curvilinear coordinate system ($\xi$,$\eta$,$\zeta$) as \citep{Jiang&Feng2010}
\begin{equation}
    \frac{\partial \mathbf{Q}}{\partial t}+\frac{\partial}{\partial \xi} \left (\frac{\mathbf{F}}{J} \right)+\frac{\partial}{\partial \eta}\left(\frac{\mathbf{G}}{J} \right)+\frac{\partial}{\partial \zeta}\left(\frac{\mathbf{H}}{J} \right)=\frac{\mathbf{S}}{J},
\end{equation}
where $\mathbf{Q}$ is the vector of the conservative variables,
\begin{equation}
  \mathbf{Q}=  \frac{1}{J} \begin{bmatrix}
    \rho  &
    \rho u &
    \rho v &
    \rho w &
    B_x &
    B_y &
    B_z &
    E
 \end{bmatrix}^\mathrm{T}.
\end{equation}
Variables $J$ and $\rho$ denote the Jacobian and density, respectively. Velocity field elements in the cartesian coordinate system are shown with $u$, $v$, and $w$. Furthermore, $B_x$, $B_y$, and $B_z$ represent the magnetic field components in the physical space, and the total energy value, $E$, is given as 
\begin{equation}
   E=\frac{p}{\gamma-1}+\rho \frac{\textbf{V} \cdot \textbf{V}}{2}+\frac{\textbf{B} \cdot \textbf{B}}{2},
\end{equation}
where vectors \textbf{V} and \textbf{B} represent velocity and magnetic field components, respectively.\\
The flux terms $\mathbf{F}$, $\mathbf{G}$, and $\mathbf{H}$, are defined as
\begin{equation}
    \mathbf{F}=\begin{bmatrix}
    \rho U\\
    \rho u U+\xi_x p+\left(\frac{{B_x}^2}{2}+\frac{{B_y}^2}{2}+\frac{{B_z}^2}{2}\right)\xi_x-B_x B_{\xi}\\
     \rho v U+\xi_y p+\left(\frac{{B_x}^2}{2}+\frac{{B_y}^2}{2}+\frac{{B_z}^2}{2}\right)\xi_y-B_y B_{\xi}\\
     \rho w U+\xi_z p+\left(\frac{{B_x}^2}{2}+\frac{{B_y}^2}{2}+\frac{{B_z}^2}{2}\right)\xi_z-B_z B_{\xi}\\
     B_x U-u B_{\xi}\\
     B_y U-v B_{\xi}\\
     B_z U-w B_{\xi}\\
     \left(E+p+\frac{\mathbf{B} \cdot \mathbf{B}}{2}\right) U- B_{\xi}\left(u B_x + v B_y +w B_z \right)
    \end{bmatrix},
\end{equation}
\begin{equation}
       \mathbf{G}=\begin{bmatrix}
    \rho V\\
    \rho u V+\eta_x p+\left(\frac{{B_x}^2}{2}+\frac{{B_y}^2}{2}+\frac{{B_z}^2}{2} \right)\eta_x-B_x B_{\eta}\\
     \rho v V+\eta_y p+\left(\frac{{B_x}^2}{2}+\frac{{B_y}^2}{2}+\frac{{B_z}^2}{2} \right)\eta_y-B_y B_{\eta}\\
     \rho w V+\eta_z p+ \left(\frac{{B_x}^2}{2}+\frac{{B_y}^2}{2}+\frac{{B_z}^2}{2} \right)\eta_z-B_z B_{\eta}\\
     B_x V-u B_{\eta}\\
     B_y V-v B_{\eta}\\
     B_z V-w B_{\eta}\\
     \left(E+p+\frac{\mathbf{B} \cdot \mathbf{B}}{2} \right) V- B_{\eta}\left(u B_x + v B_y +w B_z \right)
    \end{bmatrix},
\end{equation}
and
\begin{equation}
           \mathbf{H}=\begin{bmatrix}
    \rho W\\
    \rho u W+\zeta_x p+\left(\frac{{B_x}^2}{2}+\frac{{B_y}^2}{2}+\frac{{B_z}^2}{2} \right)\zeta_x-B_x B_{\zeta}\\
     \rho v W+\zeta_y p+\left(\frac{{B_x}^2}{2}+\frac{{B_y}^2}{2}+\frac{{B_z}^2}{2}\right)\zeta_y-B_y B_{\zeta}\\
     \rho w W+\zeta_z p+\left(\frac{{B_x}^2}{2}+\frac{{B_y}^2}{2}+\frac{{B_z}^2}{2} \right)\zeta_z-B_z B_{\zeta}\\
     B_x W-u B_{\zeta}\\
     B_y W-v B_{\zeta}\\
     B_z W-w B_{\zeta}\\
     \left(E+p+\frac{\mathbf{B} \cdot \mathbf{B}}{2}\right) W- B_{\zeta}\left(u B_x + v B_y +w B_z \right)
    \end{bmatrix},
\end{equation}
with variables $U$, $V$, and $W$ denoting the velocity components in the computational space,
\begin{equation}
    U=u \xi_x + v \xi_y + w \xi_z, \        \ V=u \eta_x + v \eta_y + w \eta_z, \        \ W=u \zeta_x + v \zeta_y + w \zeta_z,
\end{equation}
where $\xi_x$, $\xi_y$, and $\xi_z$ are curvilinear coordinate metrics.
Additionally, $B_{\xi}$, $B_{\eta}$, and $B_{\zeta}$ variables are the transformed magnetic field components in the computational space. These variables are defined in the following equations:
\begin{equation}
     B_{\xi}=B_x \xi_x + B_y \xi_y + B_z \xi_z, \        \ B_{\eta}=B_x \eta_x + B_y \eta_y + B_z \eta_z, \        \ B_{\zeta}=B_x \zeta_x + B_y \zeta_y + B_z \zeta_z.
\end{equation}
Finally, the source term flux vector, $\mathbf{S}$, reads as
  \begin{equation}
      \mathbf{S}=\begin{bmatrix}
      0 \\
      -\left(\frac{\partial B_{\xi}}{\partial \xi} + \frac{\partial B_{\eta}}{\partial \eta} + \frac{\partial B_{\zeta}}{\partial \zeta} \right) B_x \\
      -\left(\frac{\partial B_{\xi}}{\partial \xi} + \frac{\partial B_{\eta}}{\partial \eta} + \frac{\partial B_{\zeta}}{\partial \zeta} \right) B_y \\
       -\left(\frac{\partial B_{\xi}}{\partial \xi} + \frac{\partial B_{\eta}}{\partial \eta} + \frac{\partial B_{\zeta}}{\partial \zeta}\right) B_z \\
       -\left(\frac{\partial B_{\xi}}{\partial \xi} + \frac{\partial B_{\eta}}{\partial \eta} + \frac{\partial B_{\zeta}}{\partial \zeta} \right) u \\
      -\left(\frac{\partial B_{\xi}}{\partial \xi} + \frac{\partial B_{\eta}}{\partial \eta} + \frac{\partial B_{\zeta}}{\partial \zeta} \right) v \\
       -\left(\frac{\partial B_{\xi}}{\partial \xi} + \frac{\partial B_{\eta}}{\partial \eta} + \frac{\partial B_{\zeta}}{\partial \zeta} \right) w \\
       -\left(\frac{\partial B_{\xi}}{\partial \xi} + \frac{\partial B_{\eta}}{\partial \eta} + \frac{\partial B_{\zeta}}{\partial \zeta}\right) \left(u B_x + v B_y + w B_z \right) 
     \end{bmatrix}.
  \end{equation}
This source term, known as Powell's source term, prevents the accumulation of the finite $\nabla \cdot \mathbf{B}$ generated by the numerical simulation at a fixed grid point by advecting the error away \citep{Hussaini&Leer1997}. This method, also known as the 8-wave formulation, works well for many problems in which $\nabla \cdot \mathbf{B}$ remains small \citep{Toth2000}. However, for problems containing strong discontinuities, this method can produce incorrect jump conditions and inaccurate results. Therefore, the common practice is to couple this method with a supplementary cleaning technique (e.g., projection scheme) to ensure that the divergence-free constraint is (approximately) satisfied \citep{Toth2000}. Since the majority of our test cases involve smooth flows and the calculated $\nabla \cdot \mathbf{B}$ is relatively small, we only apply the 8-wave formulation scheme for the divergence cleaning.
\par
It is worth mentioning that the derived characteristic boundary scheme in this study is not only limited to the 8-wave formulation since the source term are neglected while deriving the MHD eigenstructure. Therefore, the proposed method can be easily extended to other MHD discretizations using the projection or constrained transport schemes for satisfying the magnetic field divergence-free condition. This is due to the fact that the calculated eigenvalues and eigenvectors for the MHD governing equations remain valid.

\section{Numerical schemes}

\subsection{Spatial discretization schemes}
Two spatial discretization methods are used in this study, namely a sixth-order central compact scheme \citep{Lele1992} and a fifth-order Weighted Essentially Non-Oscillatory (WENO) scheme \citep{Shu1997,Jiang&Wu1999}. The former is a cost-effective, fast, and accurate finite difference scheme with spectral-like resolution that is widely used for problems containing smooth flows. However, this scheme is not suitable for problems with discontinuities, while the WENO scheme can accurately capture these discontinuities, maintaining a high order of accuracy, but at a higher computational cost.

\subsubsection{Central compact scheme}
The sixth-order compact scheme used in this work involves solving a tridiagonal system of equations for a variable's derivative. This system is given by
\begin{equation}
    \alpha \left(\frac{\partial f}{\partial \xi}\right)_{i-1}+\left(\frac{\partial f}{\partial \xi}\right)_{i}+\alpha \left(\frac{\partial f}{\partial \xi} \right)_{i+1}=a\frac{f_{i+1}-f_{i-1}}{2 \Delta \xi}+b\frac{f_{i+2}-f_{i-2}}{4 \Delta \xi},
\end{equation}
where $\alpha = 1/3$, $a=14/9$, and $b=1/9$. \par
For the boundary nodes, $i=1$, $N$ and $i=2$, $N-1$, the third-order and fourth-order one-sided compact schemes were used, respectively. This scheme has sixth-order accuracy for interior nodes and third- and fourth-order accuracy for boundary nodes. The central compact scheme can maintain a global fourth-order accuracy. This tridiagonal system is solved by using the Tridiagonal Matrix Algorithm (TDMA) \citep{Conte&Boor1980}.

\subsubsection{WENO scheme}

A finite difference version of the WENO scheme \citep{Shu1997} is implemented in our MHD solver code. The main idea of the fifth-order WENO scheme is to use a proper weighted combination of three sub-stencils for calculating the flux values. This combination is chosen such that small weight values are assigned to the stencils with discontinuities to prevent any numerical instability \citep{Jiang&Wu1999}. A finite volume discretization of the general one-dimensional scalar equation, $u_t=-f(u)_x$, can be written as
 \begin{equation}\label{eq:WENO1}
      \frac{\mathrm{d}u_i(t)}{\mathrm{d}t}=-\frac{1}{\Delta x}(\hat{f}_{i+\frac{1}{2}}-\hat{f}_{i-\frac{1}{2}}),
 \end{equation}
where $u_i$ shows the value of $u$ at $x_i$, and $\hat{f}$ is a convex flux which should be constructed in a way to prevent any numerical instabilities or entropy violating conditions. As a result, the upwinding and flux splitting approaches should be considered while calculating numerical fluxes \citep{Shu1997}. Here, Lax-Friedrichs splitting \citep{Shu1997} is used, giving the positive and negative flux components as $f^{\pm}(u)=\frac{1}{2}(f(u) \pm \alpha u)$, where $\alpha$ denotes the maximum value of $|f^{'}(u)|$. After finding $f^{\pm}(u)$, we can find the desired positive and negative numerical fluxes according to the WENO scheme as below \citep{Jiang&Wu1999}:
 \begin{equation}\label{eq:WENO2}
     \hat{f}^+_{i+\frac{1}{2}}=\frac{1}{12}\left(-{f}^+_{i-1}+7{f}^+_{i}+7{f}^+_{i+1}-{f}^+_{i+2}\right)-\varphi \left(\Delta {f}^+_{i-\frac{3}{2}},\Delta {f}^+_{i-\frac{1}{2}},\Delta {f}^+_{i+\frac{1}{2}},\Delta {f}^+_{i+\frac{3}{2}}\right),
 \end{equation}
 and
 \begin{equation}\label{eq:WENO3}
     \hat{f}^-_{i+\frac{1}{2}}=\frac{1}{12}\left(-{f}^-_{i-1}+7{f}^-_{i}+7{f}^-_{i+1}-{f}^-_{i+2}\right)+\varphi \left(\Delta {f}^-_{i+\frac{5}{2}},\Delta {f}^-_{i+\frac{3}{2}},\Delta {f}^-_{i+\frac{1}{2}},\Delta {f}^-_{i-\frac{1}{2}}\right),
 \end{equation}
where, $\Delta f^{\pm}_{k+\frac{1}{2}}=f^{\pm}_{k+1}-f^{\pm}_k$ and $\varphi$ is a nonlinear function given by \citep{Jiang&Wu1999}
  \begin{equation}\label{eq:WENO4}
      \varphi(m,n,p,q)=\frac{1}{3}w_0\left(m-2n+p \right)+\frac{1}{6}\left(w_2-\frac{1}{2}\right)\left(n-2p+q\right).
  \end{equation}
Variables $w_0$ and $w_2$ are weight values for the first and third stencils. Jiang and Shu \citep{Jiang&Shu1996} suggested the following weights:
\begin{equation}\label{eq:WENO5}
    w_0=\frac{\alpha_0}{\alpha_0+\alpha_1+\alpha_2}, \        \  w_2=\frac{\alpha_2}{\alpha_0+\alpha_1+\alpha_2},
\end{equation}
 where
  \begin{equation}\label{eq:WENO6}
      \alpha_0=\frac{1}{(\epsilon+\beta_0)^2},  \        \  \alpha_1=\frac{6}{(\epsilon+\beta_1)^2}, \        \  \alpha_2=\frac{3}{(\epsilon+\beta_2)^2}.
  \end{equation}
 Here parameter $\epsilon$ is used to prevent the denominators to become zero. The value of $\epsilon$ is typically chosen to be between $10^{-5}$ and $10^{-7}$. We used a value of $\epsilon=10^{-6}$.
 The smooth indicator parameters, denoted by $\beta$, are given as 
 \begin{equation}\label{eq:WENO7}
     \beta_0=13(m-n)^2+3(m-3n)^2,  \        \  \beta_1=13(n-p)^2+3(n+p)^2, \        \  \beta_2=13(p-q)^2+3(3p-q)^2.
\end{equation}
Combining Eqs. (\ref{eq:WENO2}) and (\ref{eq:WENO3}) results in
\begin{equation}\label{eq:WENO8}
    \hat{f}_{i+\frac{1}{2}} = \frac{1}{12}\left(-f_{i-1}+7{f}_{i}+7{f}_{i+1}-{f}_{i+2} \right)-\varphi \left(\Delta {f}^+_{i-\frac{3}{2}},\Delta {f}^+_{i-\frac{1}{2}},\Delta {f}^+_{i+\frac{1}{2}},\Delta {f}^+_{i+\frac{3}{2}} \right) +\varphi \left(\Delta {f}^-_{i+\frac{5}{2}},\Delta {f}^-_{i+\frac{3}{2}},\Delta {f}^-_{i+\frac{1}{2}},\Delta {f}^-_{i-\frac{1}{2}} \right).
\end{equation}
Similarly, $\hat{f}_{i-\frac{1}{2}}$ can be calculated, and by substituting $\hat{f}_{i \pm \frac{1}{2}}$ values in Eq. (\ref{eq:WENO1}), we can calculate the derivative of $u$ at $x_i$.
\par
There are several ways to generalize the scalar WENO scheme to the system of equations \citep{Shu1997}. The most straightforward approach, implemented in this study, is a component-wise finite difference method \citep{Shu1997}. The component by component version of the WENO scheme is simple, cost-effective, and reasonably robust for many problems. In this method, the flux splitting procedure is applied separately to each component of $u$ to construct the numerical flux values ${\hat{f}}_{i+\frac{1}{2}}$. Afterward, the spatial derivative is calculated using Eq. (\ref{eq:WENO1}). Furthermore, the same procedure should be applied in each direction for multi-dimensional problems.

\subsection{Temporal integration scheme}\label{sec:RK}

Consider a system of differential equations given by $\frac{\partial \mathbf{U}}{\partial t}=\mathbf{F}(\mathbf{U},t)$, where $\mathbf{U}$ is the vector of conservative variables and $\mathbf{F}$ is a right-hand-side operator. For a known solution, $\mathbf{U}^n$, the low-storage format of the fourth-order Runge-Kutta scheme \citep{Mitchell1995} is implemented in our in-house code to approximate the solution at the next time step, $\mathbf{U}^{n+1}$. This method uses the following four steps to approximate the solution at the time step $n+1$:
\begin{enumerate}
    \item \text{The first step (Euler Predictor): }{ \begin{subequations}
    \begin{align}
          \mathbf{U}^{'} = \mathbf{F}(\mathbf{U}^n,t) \\
        \mathbf{Q}_1=\mathbf{U}^n + \frac{\Delta t}{2}\mathbf{U}^{'} \\
        \mathbf{Q}_2=\mathbf{U}^n + \frac{\Delta t}{6}\mathbf{U}^{'}
    \end{align}
     \end{subequations}}
    \item \text{The second step (Euler Corrector): }{ \begin{subequations}
    \begin{align}
          \mathbf{U}^{'} = \mathbf{F}\left(\mathbf{Q}_1,t+\frac{\Delta t}{2}\right) \\
        \mathbf{Q}_1=\mathbf{U}^n + \frac{\Delta t}{2}\mathbf{U}^{'} \\
        \mathbf{Q}_2=\mathbf{Q}_2 + \frac{\Delta t}{3}\mathbf{U}^{'}
    \end{align}
     \end{subequations}}
    \item \text{The third step (Leapfrog Predictor): }{ \begin{subequations}
    \begin{align}
          \mathbf{U}^{'} = \mathbf{F}\left(\mathbf{Q}_1,t+\frac{\Delta t}{2}\right) \\
        \mathbf{Q}_1=\mathbf{U}^n + \Delta t \mathbf{U}^{'} \\
        \mathbf{Q}_2=\mathbf{Q}_2 + \frac{\Delta t}{3}\mathbf{U}^{'}
    \end{align}
     \end{subequations}}
    \item \text{The fourth step (Milne Corrector) : }{ \begin{subequations}
    \begin{align}
          \mathbf{U}^{'} = \mathbf{F}\left(\mathbf{Q}_1,t+\Delta t \right) \\
        \mathbf{U}^{n+1}=\mathbf{Q}_2+\frac{\Delta t}{6} \mathbf{U}^{'}
    \end{align}
     \end{subequations}}
\end{enumerate}
The convergence study of the implemented MHD solver is provided in Appendix A for interested readers.
 
\section{Characteristic waves and boundary conditions}
In this section, the procedure of implementing the characteristic boundary scheme for a one-dimensional system of governing equations is presented. The same methodology will be extended to the MHD equations in three-dimensional curvilinear coordinates using the complete derived set of MHD eigenstructure provided in Sec. \ref{sec:MHDeigenstructure}. In Sec. \ref{sec:BC}, we delineate how different boundary conditions, namely non-reflecting inlet/outlet, solid wall, and constant pressure, can be imposed by employing a characteristic boundary scheme.

\subsection{Characteristic boundary scheme}
The characteristic boundary scheme specifies the desired boundary conditions by identifying and manipulating the different waves crossing the computational boundaries \citep{Poinsot&Veynante2005}. The information from the outgoing waves are used to satisfy the desired boundary conditions by imposing proper constraints on the incoming waves \citep{Cimino&Krause2016}.
\par
\begin{figure}[!t]
\centering
\includegraphics[scale= 0.4]{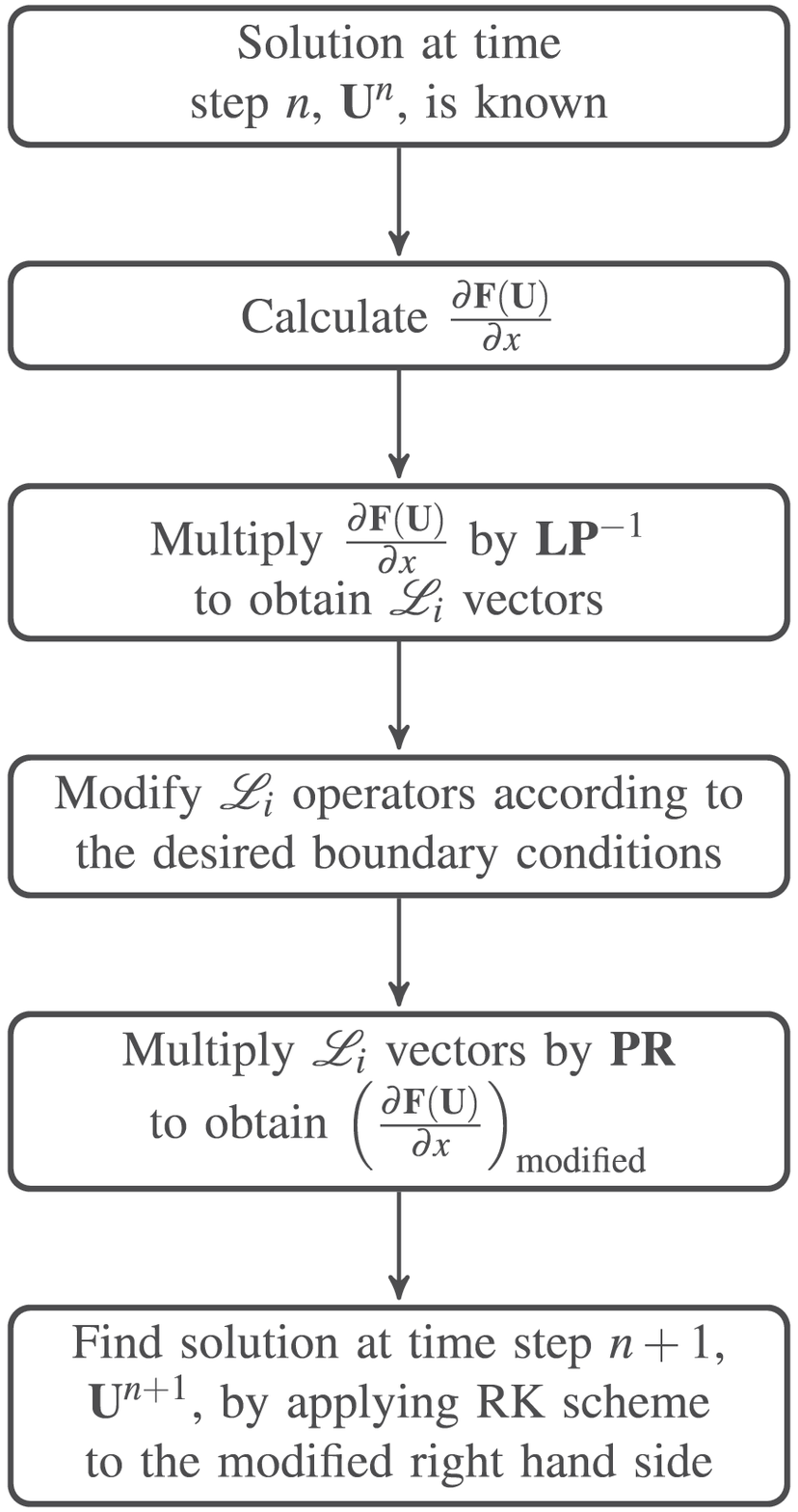}
\caption{A flowchart for implementing the characteristic boundary scheme for the general system of governing equations.}
\label{fig:CharBCMethod}
\end{figure}

Consider the conservative form of the one-dimensional system of governing equations in the cartesian coordinate system
\begin{equation}\label{eq:charBC1}
 \frac{\partial \mathbf{U}}{\partial t}+\frac{\partial \mathbf{F}(\mathbf{U})}{\partial x}=\textbf{0},
\end{equation}
where $\mathbf{U}$ is the vector of conservative variables and $\mathbf{F}$ is the flux vector. Eq. (\ref{eq:charBC1}) describes the relationship between the temporal rate of the conservative field over an infinitesimal control volume and the flux of that field crossing the control volume boundaries \citep{Thompson1987}. Eq. (\ref{eq:charBC1}) can also be written in primitive form as \citep{Thompson1987}:
\begin{equation}\label{eq:charBC2}
    \frac{\partial \mathbf{V}}{\partial t}+\mathbf{A} \frac{\partial \mathbf{V}}{\partial x}=\textbf{0},
\end{equation}
where $\mathbf{V}$ is the vector of primitive variables and $\mathbf{A}$ is known as the Jacobian matrix of the primitive variables in the $x$-direction.
By defining $\mathbf{P}=\partial \mathbf{U}/\partial \mathbf{V}$ and $\mathbf{Q}=\partial \mathbf{F}/\partial \mathbf{V}$, we can rewrite Eq. (\ref{eq:charBC1}) as \citep{Poinsot&Lele1992,Thompson1987,Hedstrom1979}:
\begin{equation}\label{eq:charBC3}
    \frac{\partial \mathbf{U}}{\partial t}+\mathbf{Q}\frac{\partial \mathbf{V}}{\partial {x}}=\textbf{0}.
\end{equation}
Multiplying Eq. (\ref{eq:charBC3}) by $\mathbf{P}^{-1}$ will result in Eq. (\ref{eq:charBC2}) in which $\mathbf{P}^{-1}\mathbf{Q}=\mathbf{A}$. Therefore, for converting the conservative form of the governing equations to the primitive form, we should multiply Eq. (\ref{eq:charBC1}) by $\mathbf{P}^{-1}$. Matrix $\mathbf{A}$ is a diagonalizable matrix and can be written in the form $\mathbf{A}=\mathbf{R}\mathbf{\Lambda}\mathbf{L}$, where $\mathbf{R}$ and $\mathbf{L}$ are the right and left eigenvectors, and $\mathbf{\Lambda}$ is the diagonal matrix of the corresponding eigenvalues. Using this definition, we can express Eq. (\ref{eq:charBC1}) as
\begin{equation}\label{eq:charBC4}
    \frac{\partial \mathbf{V}}{\partial t}+\mathbf{R}\mathbf{\Lambda}\mathbf{L} \frac{\partial \mathbf{V}}{\partial x}=\textbf{0}.
\end{equation}
By multiplying both sides of Eq. (\ref{eq:charBC4}) by the left eigenvector and noting that $\mathbf{LR}=\mathbf{I}$ (identity matrix), we obtain
\begin{equation}\label{eq:charBC5}
     \mathbf{L} \frac{\partial \mathbf{V}}{\partial t}+\Lambda \mathbf{L} \frac{\partial \mathbf{V}}{\partial x}=\textbf{0}, \          \ \text{or} \          \  \mathbf{l}_i \frac{\partial \mathbf{V}}{\partial t}+\lambda_i \mathbf{l}_i \frac{\partial \mathbf{V}}{\partial x}=\textbf{0},
\end{equation}
where $\mathbf{l}_i$ denotes the $i$-th row of the left eigenvector and $\lambda_i$ is its corresponding eigenvalue. Equation (\ref{eq:charBC5}) shows that by projecting the primitive variables in the eigenvector space, we can decouple the governing equations which can be treated individually. For simplicity, we shall denote $\lambda_i \mathbf{l}_i \frac{\partial \mathbf{V}}{\partial x}$ by operators $\mathcal{L}_i$ hereafter. In order to express the derivative of the primitive variables as a function of $\mathcal{L}_i$ operators, we multiply Eq. (\ref{eq:charBC5}) by the right eigenvectors. This operation will result in 
\begin{equation}\label{eq:charBC6}
    \frac{\partial \mathbf{V}}{\partial t}+\mathbf{R}\mathcal{L}=\textbf{0},  \          \ \text{or} \          \ \frac{\partial \mathbf{V}}{\partial t}+\mathbf{r}_i \mathcal{L}_i = \textbf{0},
\end{equation}
where $\mathbf{r}_i$ denotes the columns of the right eigenvector. At this step, the time evolution of the primitive variables at the boundaries is derived as a function of $\mathcal{L}_i$ operators. Using this set of equations, we can impose the desired boundary conditions by modifying $\mathcal{L}_i$ operators accordingly, i.e., $\mathbf{L} \,\partial \mathbf{V}/\partial t+ \mathcal{L}_{\text{modified}} =\textbf{0}$. We can convert this modified equation to the conservative form by multiplying it by the right eigenvector and $\mathbf{P}$, respectively. This operation will result in $\partial \mathbf{U}/\partial t+\left(\partial \mathbf{F}(\mathbf{U})/\partial x\right)_{\text{modified}}=\textbf{0}$, which implements the desired boundary condition by modifying the right-hand-side operator of the Runge-Kutta scheme (see section \ref{sec:RK}). Figure \ref{fig:CharBCMethod} shows the procedure for implementing the characteristic boundary scheme for the system of governing equations.
\par
It is worth noting that the described procedure of implementing the characteristic boundary condition is stable in discretization. This is due to the fact that the characteristic boundary treatment is done after the flux discretization. Therefore, it would not create any instability in the numerical discretization.

\subsection{MHD eigenstructure in the general curvilinear coordinate system}\label{sec:MHDeigenstructure}
Consider a three-dimensional set of MHD equations written in the general form of
\begin{equation}\label{eq:MHDeigen1}
    \frac{\partial \mathbf{V}}{\partial t} + \mathbf{A}_{\xi} \frac{\partial \mathbf{V}}{\partial \xi}+\mathbf{A}_{\eta} \frac{\partial \mathbf{V}}{\partial \eta}+\mathbf{A}_{\zeta}\frac{\partial \mathbf{V}}{\partial \zeta}=\textbf{0}.
\end{equation}
 Matrices $\mathbf{A}_{\xi}$, $\mathbf{A}_{\eta}$, and $\mathbf{A}_{\zeta}$ are Jacobians of the primitive variables in $\xi$, $\eta$, and $\zeta$ directions in the general curvilinear coordinate system. 
 \par
 The matrix $\mathbf{A}_{\xi}$ is defined as
\begin{equation}\label{eq:MHDeigen2}
    \mathbf{A}_{\xi}=   
    \setlength{\arraycolsep}{1.5pt}
   \renewcommand{\arraystretch}{0.9} 
    \begin{bmatrix}
    U & \rho \xi_x & \rho \xi_y & \rho \xi_z & 0 & 0 & 0 & 0 \\
     0 & U & 0 & 0 & -\frac{B_y}{\rho} \xi_y-\frac{B_z}{\rho} \xi_z& \frac{B_y}{\rho} \xi_x & \frac{B_z}{\rho} \xi_x & \frac{\xi_x}{\rho} \\
     0 & 0 & U & 0 & \frac{B_x}{\rho} \xi_y & -\frac{B_x}{\rho} \xi_x-\frac{B_z}{\rho} \xi_z & \frac{B_z}{\rho} \xi_y &  \frac{\xi_y}{\rho} \\
      0 & 0 & 0 & U & \frac{B_x}{\rho}\xi_z & \frac{B_y}{\rho} \xi_z & -\frac{B_x}{\rho}\xi_x - \frac{B_y}{\rho} \xi_y & \frac{\xi_z }{\rho} \\
      0 & -B_y \xi_y-B_z \xi_z & B_x \xi_y & B_x \xi_z & U & 0 & 0 & 0 \\
      0 & B_y \xi_x & -B_x \xi_x-B_z \xi_z & B_y \xi_z & 0 & U & 0 & 0 \\
      0 & B_z \xi_x & B_z \xi_y & -B_x \xi_x -B_y \xi_y & 0 & 0 & U & 0 \\
      0 & \gamma p \xi_x & \gamma p \xi_y & \gamma p \xi_z & 0 & 0 & 0 & U 
    \end{bmatrix},
\end{equation}
where $ U=u \xi_x+ v \xi_y + w \xi_z$. Matrices $\mathbf{A}_\eta$ and $\mathbf{A}_\zeta$ can be defined akin to Eq. (\ref{eq:MHDeigen2}) by replacing the variable $\xi$ with $\eta$ and $\zeta$, respectively. The eigenvalues of the matrix $\mathbf{A}_{\xi}$ are 
\begin{align}\label{eq:MHDeigen3}
\begin{aligned}
{\lambda_1}=U-{c_{\mathrm{f}}}, \     \ {\lambda_2}=U-{c_{\mathrm{a}}}, \     \ {\lambda_3}=U-{c_{\mathrm{s}}}, \     \ {\lambda_4}=U,
\\
\qquad {\lambda_5}=U, \     \ {\lambda_6}=U+{c_{\mathrm{s}}}, \     \ {\lambda_7}=U+{c_{\mathrm{a}}}, \   \text{and} \ {\lambda_8}=U+{c_{\mathrm{f}}},
\end{aligned}
\end{align}
where $c_{\mathrm{f}}$ and $c_{\mathrm{s}}$ are the fast and slow magnetosonic waves derived in the general curvilinear coordinate system. These variables are defined as
\begin{equation}\label{eq:MHDeigen4}
    {c_{\mathrm{f}}}^2=\frac{1}{2}\left[\left({{a}}^2+ \left|{{b}}^2 \right| \right)+\sqrt{\left[{\left({a}^2+\left|{{b}}^2 \right|\right)^2-4 {{a}}^{2} {{b_{\xi}}}^2}\right]}\right], \        \ {c_{\mathrm{s}}}^2=\frac{1}{2}\left[\left({{a}}^2+ \left|{{b}}^2 \right| \right)-\sqrt{\left[{\left({a}^2+ \left|{{b}}^2 \right| \right)^2-4 {{a}}^{2} {{b_{\xi}}}^2} \right]} \right],
\end{equation}
where 
\begin{equation}\label{eq:MHDeigen5}
   {{a}}^{2}=\frac{\gamma p}{\rho}\left({\xi_x}^2 + {\xi_y}^2 + {\xi_z}^2 \right), \        \  \left|{{b}}^2 \right|=\left({b_x}^2+{b_y}^2 + {b_z}^2 \right)\left({\xi_x} ^2 + {\xi_y}^2+ {\xi_z} ^2 \right), \        \ {b_{\xi}}=b_x \xi_x + b_y \xi_y + b_z \xi_z, \        \ \mathbf{b}=\mathbf{B}/\sqrt\rho.
\end{equation}
 Variable ${c_{\mathrm{a}}}$ is the Alfv\'en wave speed in the curvilinear coordinates, which is given by
\begin{equation}\label{eq:MHDeigen6}
    {c_{\mathrm{a}}}={{b_{\xi}}}.
\end{equation} 
Eigenvalues ${\lambda_1}$ and ${\lambda_8}$ are the propagation speed of the fast magnetosonic characteristic waves, while eigenvalues ${\lambda_3}$ and ${\lambda_6}$ are related to the slow magnetosonic waves. Eigenvalues ${\lambda_2}$ and ${\lambda_7}$ correspond to the Alfv\'en characteristic waves. There are two equal eigenvalues ${\lambda_4} = {\lambda_5} = u$ in Eq. (\ref{eq:MHDeigen3}). The former is the propagation speed of the entropy wave, whereas the latter corresponds to a fictitious wave, convecting $\nabla \cdot \mathbf{B}$ error. \par
Magnetosonic waves (slow and fast) are analogous to the acoustic ones that are modified by the presence of the magnetic field. These waves are longitudinal and compressible. The restoring force consists of the magnetic pressure, $B_i B_i$, and the plasma pressure, $p$. In the case of the slow magnetosonic mode, density and the absolute value of the magnetic field perturbations are out of phase, while they are in phase for the fast magnetosonic modes. Alfv\'en waves propagate in a direction parallel to the magnetic field. The magnetic tension force, $B_i B_j$, is known as a restoring force in this case. Alfv\'en waves do not affect the density field and hence are considered incompressible.
\\
The right eigenvector of $\mathbf{A}_{\xi}$ is
\begin{equation}\label{eq:MHDeigen7}
\begin{aligned}
&{r_1}=\begin{array}{ccccc}
 \bigl[\rho {\alpha_{\mathrm{f}}} & \beta_x c_{\mathrm{s}} \alpha_{\mathrm{s}} S-\frac{\xi_x c_{\mathrm{f}} \alpha_{\mathrm{s}} a}{b_{\perp} {h_{\xi}}^{2}} & \beta_y c_{\mathrm{s}} \alpha_{\mathrm{s}} S-\frac{\xi_y c_{\mathrm{f}} \alpha_{\mathrm{s}} a}{b_{\perp} {h_{\xi}}^{2}}
 & \beta_z c_{\mathrm{s}} \alpha_{\mathrm{s}} S-\frac{\xi_z c_{\mathrm{f}} \alpha_{\mathrm{s}} a}{b_{\perp} {h_{\xi}}^{2}}  & \beta_x a \alpha_{\mathrm{s}} \sqrt{\rho} -\frac{\xi_x \beta_{\xi} a \alpha_{\mathrm{s}} \sqrt{\rho}}{{h_{\xi}}^{2}} \\
& \beta_y a \alpha_{\mathrm{s}} \sqrt{\rho} -\frac{\xi_y \beta_{\xi} a \alpha_{\mathrm{s}} \sqrt{\rho}}{{h_{\xi}}^{2}} & \beta_z a \alpha_{\mathrm{s}} \sqrt{\rho} -\frac{\xi_z \beta_{\xi} a \alpha_{\mathrm{s}} \sqrt{\rho}}{{h_{\xi}}^{2}} & 
\frac{\rho a^2 \alpha_{\mathrm{f}}}{{h_{\xi}}^{2}}  \bigr],
 \end{array} \\
 &{r_2}=\begin{array}{ccccc}
 \bigl[0 & \beta_z \xi_y S-\beta_y \xi_z S & \beta_x \xi_z S-\beta_z \xi_x S & \beta_y \xi_x S-\beta_x \xi_y S  & 
 \sqrt{\rho}\beta_z \xi_y -\sqrt{\rho}\beta_y \xi_z \\ &  \sqrt{\rho}\beta_x \xi_z -\sqrt{\rho}\beta_z \xi_x & \sqrt{\rho}\beta_y \xi_x -\sqrt{\rho}\beta_x \xi_y & 0 \bigr],
 \end{array}\\
 &{r_3}=\begin{array}{ccccc}
\bigl[\rho {\alpha_{\mathrm{s}}} & -\beta_x c_{\mathrm{f}} \alpha_{\mathrm{f}} S+\frac{\xi_x c_{\mathrm{s}} \alpha_{\mathrm{f}} a}{b_{\perp} {h_{\xi}}^{2}} & -\beta_y c_{\mathrm{f}} \alpha_{\mathrm{f}} S+\frac{\xi_y c_{\mathrm{s}} \alpha_{\mathrm{f}} a}{b_{\perp} {h_{\xi}}^{2}}
 & -\beta_z c_{\mathrm{f}} \alpha_{\mathrm{f}} S+\frac{\xi_z c_{\mathrm{s}} \alpha_{\mathrm{f}} a}{b_{\perp} {h_{\xi}}^{2}}  & -\beta_x a \alpha_{\mathrm{f}} \sqrt{\rho} +\frac{\xi_x \beta_{\xi} a \alpha_{\mathrm{f}} \sqrt{\rho}}{{h_{\xi}}^{2}} \\
 & -\beta_y a \alpha_{\mathrm{f}} \sqrt{\rho} +\frac{\xi_y \beta_{\xi} a \alpha_{\mathrm{f}} \sqrt{\rho}}{{h_{\xi}}^{2}} & -\beta_z a \alpha_{\mathrm{f}} \sqrt{\rho} +\frac{\xi_z \beta_{\xi} a \alpha_{\mathrm{f}} \sqrt{\rho}}{{h_{\xi}}^{2}} & 
 \frac{\rho a^2 \alpha_{\mathrm{s}}}{{h_{\xi}}^{2}}  \bigr],
 \end{array}\\
 &{r_4}=\begin{bmatrix}
1 & 0 & 0 & 0 & 0 & 0 & 0 & 0
  \end{bmatrix},\\
    &{r_5}=\begin{bmatrix}
0 & 0 & 0 & 0& \xi_x & \xi_y & \xi_z & 0
  \end{bmatrix},\\
&{r_6}=\begin{array}{ccccc}
 \bigl[\rho {\alpha_{\mathrm{s}}} & \beta_x c_{\mathrm{f}} \alpha_{\mathrm{f}} S-\frac{\xi_x c_{\mathrm{s}} \alpha_{\mathrm{f}} a}{b_{\perp} {h_{\xi}}^{2}} & \beta_y c_{\mathrm{f}} \alpha_{\mathrm{f}} S-\frac{\xi_y c_{\mathrm{s}} \alpha_{\mathrm{f}} a}{b_{\perp} {h_{\xi}}^{2}}
 & \beta_z c_{\mathrm{f}} \alpha_{\mathrm{f}} S-\frac{\xi_z c_{\mathrm{s}} \alpha_{\mathrm{f}} a}{b_{\perp} {h_{\xi}}^{2}}  & -\beta_x a \alpha_{\mathrm{f}} \sqrt{\rho} +\frac{\xi_x \beta_{\xi} a \alpha_{\mathrm{f}} \sqrt{\rho}}{{h_{\xi}}^{2}} \\
 & -\beta_y a \alpha_{\mathrm{f}} \sqrt{\rho} +\frac{\xi_y \beta_{\xi} a \alpha_{\mathrm{f}} \sqrt{\rho}}{{h_{\xi}}^{2}} & -\beta_z a \alpha_{\mathrm{f}} \sqrt{\rho} +\frac{\xi_z \beta_{\xi} a \alpha_{\mathrm{f}} \sqrt{\rho}}{{h_{\xi}}^{2}} & 
  \frac{\rho a^2 \alpha_{\mathrm{s}}}{{h_{\xi}}^{2}}  \bigr],
 \end{array}\\
  &{r_7}=\begin{array}{ccccc}
  \bigl[0 & -\beta_z \xi_y S+\beta_y \xi_z S & -\beta_x \xi_z S+\beta_z \xi_x S & -\beta_y \xi_x S+\beta_x \xi_y S & 
 \sqrt{\rho}\beta_z \xi_y -\sqrt{\rho}\beta_y \xi_z \\ &  \sqrt{\rho}\beta_x \xi_z -\sqrt{\rho}\beta_z \xi_x & \sqrt{\rho}\beta_y \xi_x -\sqrt{\rho}\beta_x \xi_y & 0 \bigr],
  \end{array}\\
  &{r_8}=\begin{array}{ccccc}
 \bigl[\rho {\alpha_{\mathrm{f}}} & -\beta_x c_{\mathrm{s}} \alpha_{\mathrm{s}} S+\frac{\xi_x c_{\mathrm{f}} \alpha_{\mathrm{s}} a}{b_{\perp} {h_{\xi}}^{2}} & -\beta_y c_{\mathrm{s}} \alpha_{\mathrm{s}} S+\frac{\xi_y c_{\mathrm{f}} \alpha_{\mathrm{s}} a}{b_{\perp} {h_{\xi}}^{2}}
 & -\beta_z c_{\mathrm{s}} \alpha_{\mathrm{s}} S+\frac{\xi_z c_{\mathrm{f}} \alpha_{\mathrm{s}} a}{b_{\perp} {h_{\xi}}^{2}} & \beta_x a \alpha_{\mathrm{s}} \sqrt{\rho} -\frac{\xi_x \beta_{\xi} a \alpha_{\mathrm{s}} \sqrt{\rho}}{{h_{\xi}}^{2}} \\
  & \beta_y a \alpha_{\mathrm{s}} \sqrt{\rho} -\frac{\xi_y \beta_{\xi} a \alpha_{\mathrm{s}} \sqrt{\rho}}{{h_{\xi}}^{2}} & \beta_z a \alpha_{\mathrm{s}} \sqrt{\rho} -\frac{\xi_z \beta_{\xi} a \alpha_{\mathrm{s}} \sqrt{\rho}}{{h_{\xi}}^{2}} & 
 \frac{\rho a^2 \alpha_{\mathrm{f}}}{{h_{\xi}}^{2}}  \bigr],
 \end{array}
\end{aligned}
\end{equation}
where ${r_1}$ to ${r_8}$ are the columns of the right eigenvector corresponding to the eigenvalues ${\lambda_1}$ to ${\lambda_8}$, respectively. Dimensionless variables $\alpha_{\mathrm{s}}$, $\alpha_{\mathrm{f}}$, $\beta$, and $S$ are defined as: 
\begin{equation}\label{eq:MHDeigen8}
    {{\alpha_{\mathrm{s}}}}^2=\frac{{{c_{\mathrm{f}}}}^2-{{a}}^2}{{{c_{\mathrm{f}}}}^2-{{c_{\mathrm{s}}}}^2}, \        \ {{\alpha_{\mathrm{f}}}}^2=\frac{{{a}}^2-{{c_{\mathrm{s}}}}^2}{{c_{\mathrm{f}}}^2-{{c_{\mathrm{s}}}}^2}, \        \ {\beta_{x (y,z)}}=\frac{{b_{x (y,z)}}}{{b_{\perp}}}, \        \ {\beta_{\xi}}=\frac{{b_{\xi}}}{{b_{\perp}}}, \        \ {S}=\mathrm{sign}({b_{\xi}}). 
\end{equation}
Moreover, variables $b_\perp$ and ${h_{\xi}}^2$ are given as
\begin{equation}\label{eq:MHDeigen9}
   {b_{\perp}}=\sqrt{{{b}}^2 -{{b_{\xi}}}^2}, \        \ {h_{\xi}}^2={\xi_x}^2 + {\xi_y}^2 + {\xi_z}^2.
\end{equation}
Here, parameter ${\alpha_{\mathrm{f}}}$ (${\alpha_{\mathrm{s}}}$) shows how close fast (slow) magnetosonic waves behave in comparison with the acoustic waves \citep{Roe&Balsara1996}. 
\\
The rows of the left eigenvector corresponding to the eigenvalues ${\lambda_1}$ to ${\lambda_8}$ are calculated as
\begin{equation}\label{eq:MHDeigen10}
\begin{aligned}
    &{l_1}=\begin{array}{ccccc}
 \bigl[0 & -\frac{c_{\mathrm{f}} \alpha_{\mathrm{s}} \xi_x}{2 a b_{\perp}}+\frac{\beta_x c_{\mathrm{s}} \alpha_{\mathrm{s}} {h_{\xi}}^{2} S}{2 a^2} & -\frac{c_{\mathrm{f}} \alpha_{\mathrm{s}} \xi_y}{2 a b_{\perp}}+\frac{\beta_y c_{\mathrm{s}} \alpha_{\mathrm{s}} {h_{\xi}}^{2} S}{2 a^2} & -\frac{c_{\mathrm{f}} \alpha_{\mathrm{s}} \xi_z}{2 a b_{\perp}}+\frac{\beta_z c_{\mathrm{s}} \alpha_{\mathrm{s}} {h_{\xi}}^{2} S}{2 a^2}
 &  \frac{\beta_x \alpha_{\mathrm{s}} {h_{\xi}}^{2}-\beta_{\xi} \alpha_{\mathrm{s}} \xi_x }{2 a \sqrt{\rho}} \\ &  \frac{\beta_y \alpha_{\mathrm{s}} {h_{\xi}}^{2}-\beta_{\xi} \alpha_{\mathrm{s}} \xi_y }{2 a \sqrt{\rho}} & \frac{\beta_z \alpha_{\mathrm{s}} {h_{\xi}}^{2}-\beta_{\xi} \alpha_{\mathrm{s}} \xi_z}{2 a \sqrt{\rho}} & \frac{\alpha_{\mathrm{f}} {h_{\xi}}^{2}}{2 \rho a^2}  \bigr],
 \end{array}\\ 
    &{l_2}=\begin{array}{ccccccc}
 \bigl[0 & \frac{\beta_z \xi_y-\beta_y \xi_z}{2} & \frac{\beta_x \xi_z-\beta_z \xi_x}{2} & \frac{\beta_y \xi_x-\beta_x \xi_y}{2} 
 & \frac{\beta_z \xi_y S-\beta_y \xi_z S}{2\sqrt{\rho}} & \frac{\beta_x \xi_z S-\beta_z \xi_x S}{2\sqrt{\rho}} \\ & \frac{\beta_y \xi_x S-\beta_x \xi_y S}{2\sqrt{\rho}} & 0 \bigr],
 \end{array}\\ 
    &{l_3}=\begin{array}{ccccc}
 \bigl[0 & \frac{c_{\mathrm{s}} \alpha_{\mathrm{f}} \xi_x}{2 a b_{\perp}}-\frac{\beta_x c_{\mathrm{f}} \alpha_{\mathrm{f}} {h_{\xi}}^{2} S}{2 a^2} & \frac{c_{\mathrm{s}} \alpha_{\mathrm{f}} \xi_y}{2 a b_{\perp}}-\frac{\beta_y c_{\mathrm{f}} \alpha_{\mathrm{f}} {h_{\xi}}^{2} S}{2 a^2} & \frac{c_{\mathrm{s}} \alpha_{\mathrm{f}} \xi_z}{2 a b_{\perp}}-\frac{\beta_z c_{\mathrm{f}} \alpha_{\mathrm{f}} {h_{\xi}}^{2} S}{2 a^2}
 &  \frac{-\beta_x \alpha_{\mathrm{f}} {h_{\xi}}^{2}+\beta_{\xi} \alpha_{\mathrm{f}} \xi_x }{2 a \sqrt{\rho}} \\ &  \frac{-\beta_y \alpha_{\mathrm{f}} {h_{\xi}}^2+\beta_{\xi} \alpha_{\mathrm{f}} \xi_y }{2 a \sqrt{\rho}} & \frac{-\beta_z \alpha_{\mathrm{f}} {h_{\xi}}^{2}+\beta_{\xi} \alpha_{\mathrm{f}} \xi_z}{2 a \sqrt{\rho}} & \frac{\alpha_{\mathrm{s}} {h_{\xi}}^{2}}{2 \rho a^2}  \bigr],
 \end{array}\\ 
  &{l_4}=\begin{bmatrix}
  1 & 0 & 0 & 0 & 0 & 0 & 0 & -\frac{ {h_{\xi}}^{2}}{{{a}}^2} 
  \end{bmatrix},\\ 
   &{l_5}=\begin{bmatrix}
    0 & 0 & 0 & 0 & \frac{\xi_x}{{h_{\xi}}^{2}} & \frac{\xi_y}{{h_{\xi}}^{2}}
    & \frac{\xi_z}{{h_{\xi}}^{2}} & 0 
  \end{bmatrix},\\ 
 &{l_6}=\begin{array}{ccccc}
 \bigl[0 & -\frac{c_{\mathrm{s}} \alpha_{\mathrm{f}} \xi_x}{2 a b_{\perp}}+\frac{\beta_x c_{\mathrm{f}} \alpha_{\mathrm{f}} {h_{\xi}}^{2} S}{2 a^2} & -\frac{c_{\mathrm{s}} \alpha_{\mathrm{f}} \xi_y}{2 a b_{\perp}}+\frac{\beta_y c_{\mathrm{f}} \alpha_{\mathrm{f}} {h_{\xi}}^{2} S}{2 a^2} & -\frac{c_{\mathrm{s}} \alpha_{\mathrm{f}} \xi_z}{2 a b_{\perp}}+\frac{\beta_z c_{\mathrm{f}} \alpha_{\mathrm{f}} {h_{\xi}}^{2} S}{2 a^2}
 &  \frac{-\beta_x \alpha_{\mathrm{f}} {h_{\xi}}^{2}+\beta_{\xi} \alpha_{\mathrm{f}} \xi_x }{2 a \sqrt{\rho}} \\ &  \frac{-\beta_y \alpha_{\mathrm{f}} {h_{\xi}}^2+\beta_{\xi} \alpha_{\mathrm{f}} \xi_y }{2 a \sqrt{\rho}} & \frac{-\beta_z \alpha_{\mathrm{f}} {h_{\xi}}^{2}+\beta_{\xi} \alpha_{\mathrm{f}} \xi_z}{2 a \sqrt{\rho}} & \frac{\alpha_{\mathrm{s}} {h_{\xi}}^{2}}{2 \rho a^2}  \bigr],
 \end{array}
\\ 
 &{l_7}=\begin{array}{ccccccc}
 \bigl[0 & \frac{-\beta_z \xi_y+\beta_y \xi_z}{2} & \frac{-\beta_x \xi_z+\beta_z \xi_x}{2} & \frac{-\beta_y \xi_x+\beta_x \xi_y}{2} 
 & \frac{\beta_z \xi_y S-\beta_y \xi_z S}{2\sqrt{\rho}} & \frac{\beta_x \xi_z S-\beta_z \xi_x S}{2\sqrt{\rho}} \\ & \frac{\beta_y \xi_x S-\beta_x \xi_y S}{2\sqrt{\rho}} & 0 \bigr],
 \end{array}\\ 
    &{l_8}=\begin{array}{ccccc}
 \bigl[0 & \frac{c_{\mathrm{f}} \alpha_{\mathrm{s}} \xi_x}{2 a b_{\perp}}-\frac{\beta_x c_{\mathrm{s}} \alpha_{\mathrm{s}} {h_{\xi}}^{2} S}{2 a^2} & \frac{c_{\mathrm{f}} \alpha_{\mathrm{s}} \xi_y}{2 a b_{\perp}}-\frac{\beta_y c_{\mathrm{s}} \alpha_{\mathrm{s}} {h_{\xi}}^{2} S}{2 a^2} & \frac{c_{\mathrm{f}} \alpha_{\mathrm{s}} \xi_z}{2 a b_{\perp}}-\frac{\beta_z c_{\mathrm{s}} \alpha_{\mathrm{s}} {h_{\xi}}^2 S}{2 a^2}
 &  \frac{\beta_x \alpha_{\mathrm{s}} {h_{\xi}}^{2}-\beta_{\xi} \alpha_{\mathrm{s}} \xi_x }{2 a \sqrt{\rho}} \\ &  \frac{\beta_y \alpha_{\mathrm{s}} {h_{\xi}}^2-\beta_{\xi} \alpha_{\mathrm{s}} \xi_y }{2 a \sqrt{\rho}} & \frac{\beta_z \alpha_{\mathrm{s}} {h_{\xi}}^{2}-\beta_{\xi} \alpha_{\mathrm{s}} \xi_z}{2 a \sqrt{\rho}} & \frac{\alpha_{\mathrm{f}} {h_{\xi}}^{2}}{2 \rho a^2}  \bigr].
 \end{array}\\
 \end{aligned}
\end{equation}
 Details for simplifying the MHD eigenstructure in the curvilinear coordinate system is given in Appendix B.
Similar sets of eigenvectors and eigenvalues can be defined for $\mathbf{A}_{\eta}$ ($\mathbf{A}_{\zeta}$), by substituting $\xi_x$, $\xi_y$, and $\xi_z$ with $\eta_x$, $\eta_y$, and $\eta_z$ ($\zeta_x$, $\zeta_y$, and $\zeta_z$), respectively. Matrices $\mathbf{P}^{-1}$ and $\mathbf{P}$ for the curvilinear coordinate case are given as \\
\begin{equation}\label{eq:MHDeigen11}
    \mathbf{P}^{-1}=
   \renewcommand{\arraystretch}{0.7} 
    \begin{bmatrix}
    J & 0 & 0 & 0 & 0 & 0 & 0 & 0 \\
    -\frac{u}{\rho}J & \frac{1}{\rho}J & 0 & 0 & 0 & 0 & 0 & 0\\
    -\frac{v}{\rho}J & 0 & \frac{1}{\rho}J & 0 & 0 & 0 & 0 & 0 \\
    -\frac{w}{\rho}J & 0 & 0 & \frac{1}{\rho}J & 0 & 0 & 0 & 0 \\
    0 & 0 & 0 & 0 & J & 0 & 0 & 0 \\
    0 & 0 & 0 & 0 & 0  & J & 0 & 0\\
    0 & 0 & 0 & 0 & 0 & 0 & J & 0\\
    \left(\frac{u^2+v^2+w^2}{2} \right)K J & -u K J & -v K J& -w K J & -B_x K J & -B_y K J & -B_z K J& K J
    \end{bmatrix},
\end{equation}
and
\begin{equation}\label{eq:MHDeigen12}
    \mathbf{P}=
    \renewcommand{\arraystretch}{0.9}
    \begin{bmatrix}
    \frac{1}{J} & 0 & 0 & 0 & 0 & 0 & 0 & 0 \\
    \frac{u}{J} & \frac{\rho}{J} & 0 & 0 & 0 & 0 & 0 & 0\\
    \frac{v}{J} & 0 & \frac{\rho}{J} & 0 & 0 & 0 & 0 & 0 \\
    \frac{w}{J} & 0 & 0 & \frac{\rho}{J} & 0 & 0 & 0 & 0 \\
    0 & 0 & 0 & 0 & \frac{1}{J} & 0 & 0 & 0 \\
    0 & 0 & 0 & 0 & 0  & \frac{1}{J} & 0 & 0\\
    0 & 0 & 0 & 0 & 0 & 0 & \frac{1}{J} & 0\\
    \left(\frac{u^2+v^2+w^2}{2J} \right) & \frac{\rho u}{J} & \frac{\rho v}{J} & \frac{\rho w}{J} & \frac{B_x}{J} &  \frac{B_y}{J} & \frac{B_z}{J} & \frac{1}{K J} 
    \end{bmatrix},\\
\end{equation}
where $K=(\gamma -1)$, and $\gamma$ is the specific heat ratio. Using the derived MHD eigenstructure in the curvilinear coordinates, we can employ the same procedure as the previous section to apply the characteristic boundary scheme.

\subsection{Implementing boundary conditions}\label{sec:BC}

\begin{figure}[!t]
\centering
\includegraphics [trim={0 3.5cm 0 0},scale=.5]{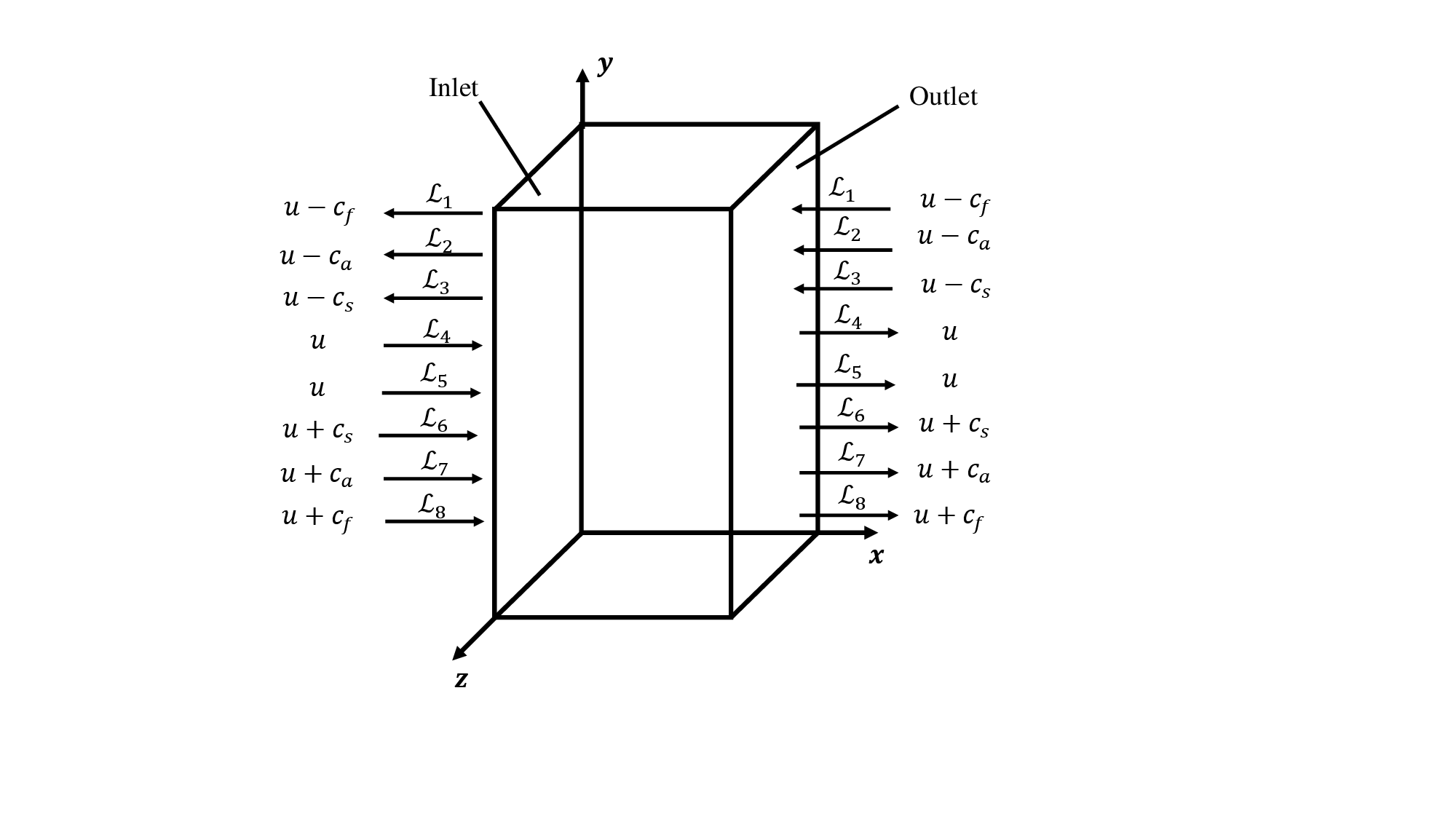}
\caption{Characteristic waves cross the boundaries in the MHD problem.}
\label{fig:MHDwaves}
\end{figure}

Characteristic equations at the boundaries contain eigenvalues of both signs, showing that characteristic waves are propagating into or out of the computational domain. Characteristic equations derived in the MHD characteristic boundary scheme allow us to study each wave separately. The outgoing waves only depend on the information received from inside the computational domain and at the boundaries. On the other hand, the incoming waves carry the information from outside the computational domain. Ignoring information from outside the computational domain leads to uncertainties in the approximation of the incoming waves.
 The number of imposed boundary conditions should be equal to the number of waves entering the computational domain. Consider the outlet boundary condition at the right boundary. Depending on the magnitude of the characteristic eigenvalues, waves corresponding to the operators $\mathcal{L}_1$ to $\mathcal{L}_3$ can be incoming or outgoing. Figure \ref{fig:MHDwaves} shows the characteristic waves at each boundary. For super magneto-slow flow, i.e., $u>c_{\mathrm{s}}$, two boundary conditions can be imposed. On the other hand, only one boundary condition can be imposed, provided the flow is super Alfv\'enic, i.e., $u>c_{\mathrm{a}}$. Finally, no boundary condition can be imposed if the flow is super magneto-fast, i.e., $u>c_{\mathrm{f}}$, because no wave enters the domain \citep{Cimino&Krause2016}. 
The time evolution of the primitive variables as a function of $\mathcal{L}_i$ operators for the one-dimensional MHD equations in the cartesian coordinates can be derived as follows
\begin{subequations}
\begin{equation}\label{eq:MHDchar1}
    \frac{\partial \rho}{\partial t}+\alpha_{\mathrm{f}} \rho \left(\mathcal{L}_1+\mathcal{L}_8 \right)+\alpha_{\mathrm{s}} \rho\left (\mathcal{L}_3+\mathcal{L}_6\right)+\mathcal{L}_4=0
\end{equation}
\begin{equation}\label{eq:MHDchar2}
    \frac{\partial u}{\partial t}+\alpha_{\mathrm{f}} c_{\mathrm{f}}\left(\mathcal{L}_8-\mathcal{L}_1 \right)+ \alpha_{\mathrm{s}} c_{\mathrm{s}} \left(\mathcal{L}_6-\mathcal{L}_{3}\right)=0,
\end{equation}
\begin{equation}\label{eq:mhdchar3}
    \frac{\partial v}{\partial t}+\alpha_{\mathrm{f}} c_{\mathrm{f}} \beta_y S \left(\mathcal{L}_6-\mathcal{L}_3 \right)+\alpha_{\mathrm{s}} c_{\mathrm{s}} \beta_y S \left(\mathcal{L}_1-\mathcal{L}_8 \right)+\beta_z \left(\mathcal{L}_7-\mathcal{L}_2 \right)=0,
\end{equation}
\begin{equation}\label{eq:MHDchar4}
     \frac{\partial w}{\partial t}+\alpha_{\mathrm{f}} c_{\mathrm{f}} \beta_z S \left(\mathcal{L}_6-\mathcal{L}_3 \right)+\alpha_{\mathrm{s}} c_{\mathrm{s}} \beta_z S \left(\mathcal{L}_1-\mathcal{L}_8 \right)+\beta_y \left(\mathcal{L}_2-\mathcal{L}_7 \right)=0,
\end{equation}
\begin{equation}\label{eq:MHDchar5}
    \frac{\partial B_x}{\partial t} + \mathcal{L}_5=0,
\end{equation}
\begin{equation}\label{eq:MHDchar6}
   \frac{\partial B_y}{\partial t}+ \alpha_{\mathrm{f}} \sqrt{\rho} \, \beta_y \, a \left(-\mathcal{L}_6-\mathcal{L}_3 \right)+\alpha_{\mathrm{s}}  \sqrt{\rho} \, \beta_y \, a \left(\mathcal{L}_1+\mathcal{L}_8 \right)-\sqrt{\rho} \, \beta_z \, S \left(\mathcal{L}_2+\mathcal{L}_7 \right)=0,
\end{equation}
\begin{equation}\label{eq:MHDchar7}
    \frac{\partial B_z}{\partial t}+\alpha_{\mathrm{f}} \sqrt{\rho} \, \beta_z \, a \left(-\mathcal{L}_6-\mathcal{L}_3 \right)+\alpha_{\mathrm{s}}  \sqrt{\rho} \, \beta_z \, a \left(\mathcal{L}_1+\mathcal{L}_8 \right)+\sqrt{\rho} \, \beta_y \, S \left(\mathcal{L}_2+\mathcal{L}_7 \right)=0,
\end{equation}
and,
\begin{equation}\label{eq:MHDchar8}
    \frac{\partial p}{\partial t} +\alpha_{\mathrm{f}} \, \rho \, a^2 \left(\mathcal{L}_1+\mathcal{L}_8 \right)+\alpha_{\mathrm{s}} \, \rho \, a^2 \left(\mathcal{L}_3+\mathcal{L}_6 \right)=0.
\end{equation}
\end{subequations}
Below, the procedure of implementing the non-reflecting inlet/outlet, solid wall, and constant pressure boundary conditions is discussed.
\subsubsection{Non-reflecting boundary condition}\label{sec:nonrefBC}
Applying non-reflective boundary conditions involves setting the magnitude of the incoming waves to zero \citep{Poinsot&Lele1992,Thompson1987,Hedstrom1979}. This guarantees that no wave can enter the computational domain. For example, consider the case of an inlet boundary condition at the left boundary. The waves associated with the operators $\mathcal{L}_4$ to $\mathcal{L}_8$ enter the domain (see Fig. \ref{fig:MHDwaves}), and their amplitude should be set to zero in order to have a non-reflecting boundary condition. Other waves related to operators $\mathcal{L}_1$ to $\mathcal{L}_3$ may be incoming or outgoing waves depending on their corresponding calculated eigenvalue signs. Non-reflecting outlet boundary conditions can also be implemented using a similar approach.
\subsubsection{Slip wall boundary condition}
For a slip wall boundary condition, the normal velocity should be set to zero at the initial condition and remain zero during the simulation. This means that the condition $\partial u/\partial t=0$ should be met at the boundary. As Eq. (\ref{eq:MHDchar2}) suggests, for imposing this condition, we set $\mathcal{L}_6$ equal to $\mathcal{L}_3$ and $\mathcal{L}_8$ equal to $\mathcal{L}_1$ or \textit{vice versa}, depending upon which one leaves or enters the computational domain. This approach guarantees that the normal velocity component remains zero and the wall boundary is satisfied.   
\subsubsection{Constant pressure outlet boundary condition}
This boundary condition implies that the pressure derivative with respect to time at the boundary should be zero throughout the simulation. Eq. (\ref{eq:MHDchar8}) suggests that the time derivative of pressure is controlled by $\mathcal{L}_1$, $\mathcal{L}_3$, $\mathcal{L}_6$, and $\mathcal{L}_8$. Suppose we have a constant pressure outlet at the right boundary. In this case, the characteristic waves corresponding to $\mathcal{L}_8$ and $\mathcal{L}_6$ are outgoing. Therefore, by setting $\mathcal{L}_1=-\mathcal{L}_8$ and $\mathcal{L}_3=-\mathcal{L}_6$, the time  derivative of the pressure remains zero at the boundary node.
\par
Using the same methodology as the above-mentioned examples, any other kind of boundary condition can be imposed. The same approach is valid for implementing different boundary conditions in the curvilinear coordinates.

\section{Numerical examples}
In this section, we first examine the derived characteristic boundary scheme for the non-magnetic test case, Sod's shock tube, to verify whether the scheme works properly for the gasdynamic case. Next, the newly designed test cases are investigated in order to study the implementation and correctness of the proposed MHD characteristic boundary scheme for the curvilinear computational domains.

\subsection{Sod shock tube test}

\begin{figure}[!t]
\centering
\includegraphics[scale=.5]{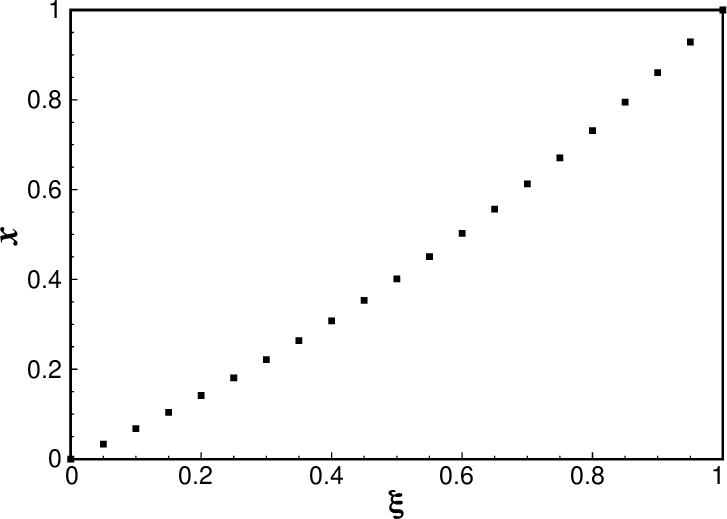}
\caption{Distribution of the physical domain points, $x$, against the computational ones, $\xi$, in the non-uniform stretched computational domain, using $20$ grid points with $\delta_x=0.8$.}
\label{fig:TestCase1_1}
\end{figure}

In this test case, the MHD numerical solver is applied to Sod's shock tube problem \citep{Sod1978} to assess if the MHD solver can correctly simulate a gasdynamic case and whether the MHD characteristic boundary scheme can be automatically reduced to the non-magnetic case. The initial conditions for the left and right regions are defined in the non-dimensional form as
\begin{equation}
            \left(\rho, u,p \right)=\begin{cases}
            (1,0,1) \text{\   \ \  \  \  \    \   \  \ \      \ \      \ \ for $x<0.5$}, \\
            (0.125,0,0.1) \        \ \text{\ \ for $x>0.5$},  
            \end{cases} 
\end{equation}
with the heat capacity ratio $\gamma = 1.4$. The fifth-order WENO and fourth-order Runge-Kutta schemes are used for the spatial discretization and time marching, respectively. The CFL number is $0.5$ and a solid wall boundary condition is applied at both ends. The computational domain consists of $600$ non-uniform stretched grid points. The size of the computational domain is $\xi \in [0, 1]$, and the grid coordinates in the physical domain are given by
 \begin{equation}\label{eq:StretchedComputationalDomain}
     x=\left(x_{\mathrm{max}}-x_{\mathrm{min}} \right)\frac{\exp{\left(\delta_x \, \xi \right)}-1}{\exp \left(\delta_x \right)-1}+x_{\mathrm{min}},
 \end{equation}
 where $\delta_x$ is the stretch factor in the $x$-direction, which is set to $\delta_x=0.8$ for this test case. Parameters $x_{\mathrm{min}}$ and $x_{\mathrm{max}}$ denote the minimum and maximum values of the physical domain, respectively, with values $x_{\mathrm{min}}=0$ and $x_{\mathrm{max}}=1$. Figure \ref{fig:TestCase1_1} shows the distribution of the physical domain points against the computational ones. It can be observed that the points are more compressed near $x=0$ and become more stretched as approaching $x=1$.
\par
This problem is simulated using two numerical solvers, the implemented MHD solver as well as the verified in-house Euler solver with the characteristic boundary scheme. In the MHD solver, all the magnetic components are set to zero, and the density, velocity, and pressure fields are initialized according to the initial conditions of the Sod shock tube problem. Setting the magnetic field to zero results in the disappearance of the slow magnetosonic and Alfv\'en waves. Furthermore, the fast magnetosonic waves are reduced to acoustic ones.
\par
\begin{figure}[!t]   
\centering
\hspace{-1.6cm} 
\subfloat[]{\includegraphics[width = 3.5in]{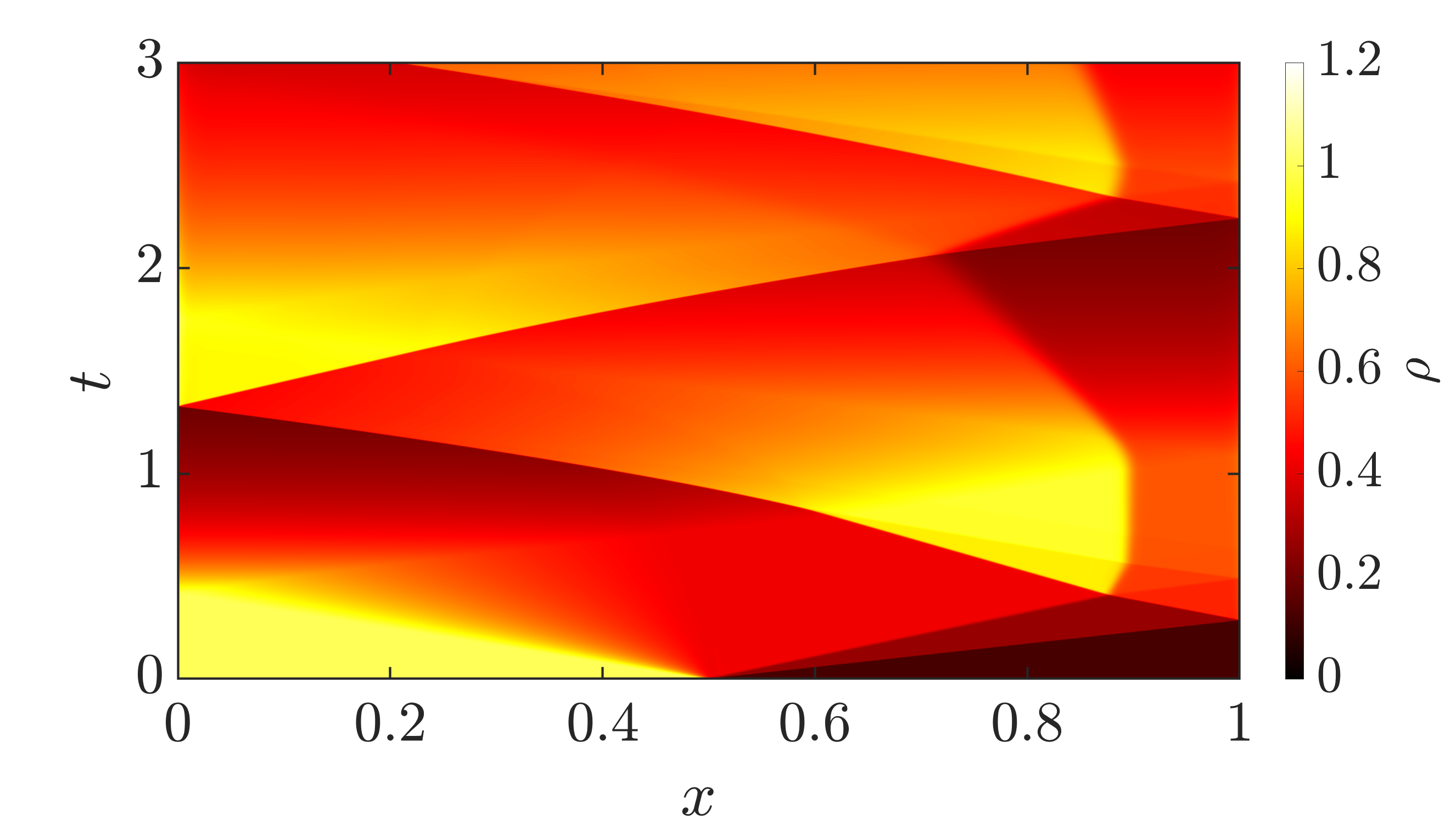}} 
\subfloat[]{\includegraphics[width = 3.5in]{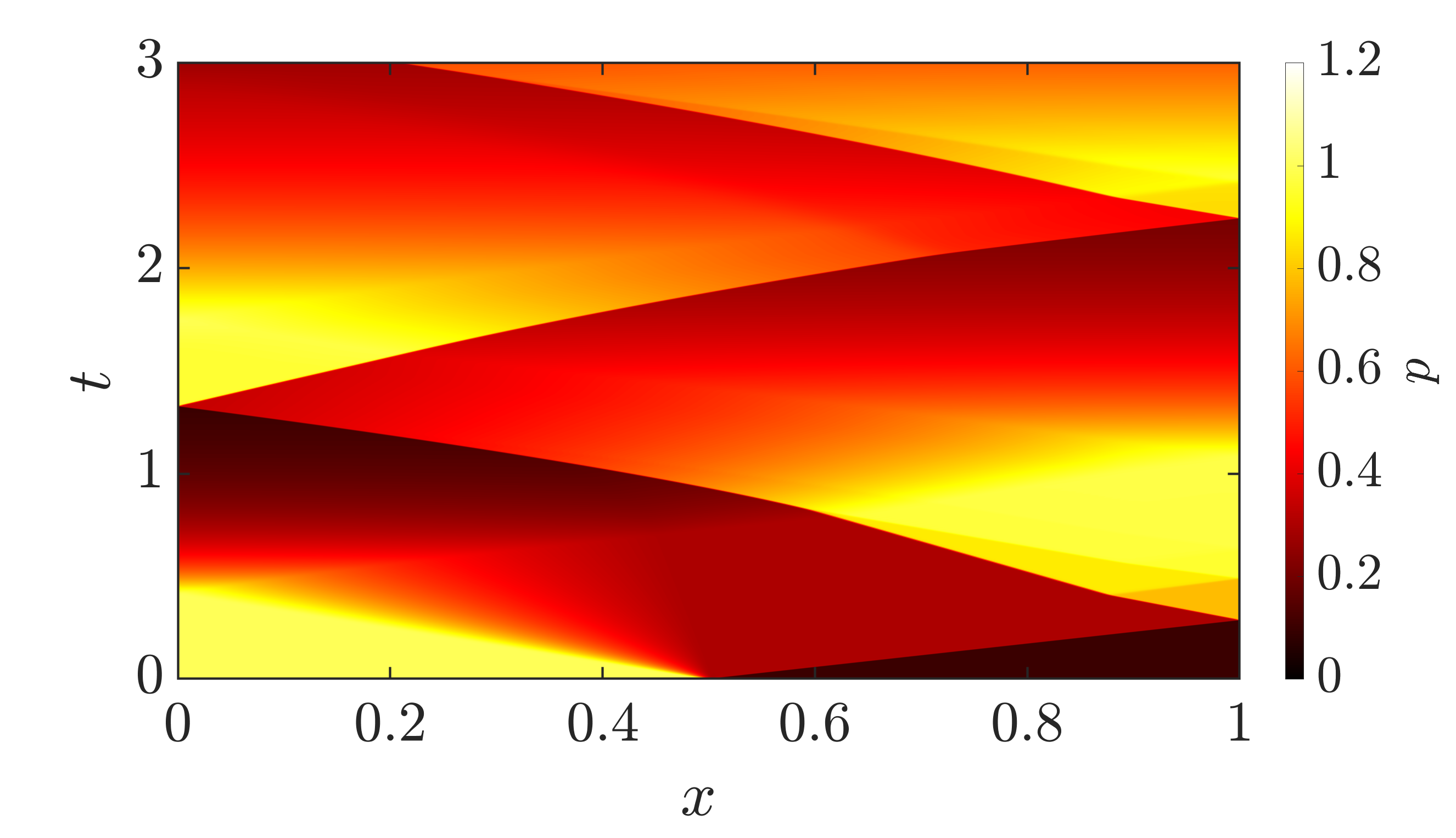}}\\
\hspace{-1.6cm}
\subfloat[]{\includegraphics[width =3.5in]{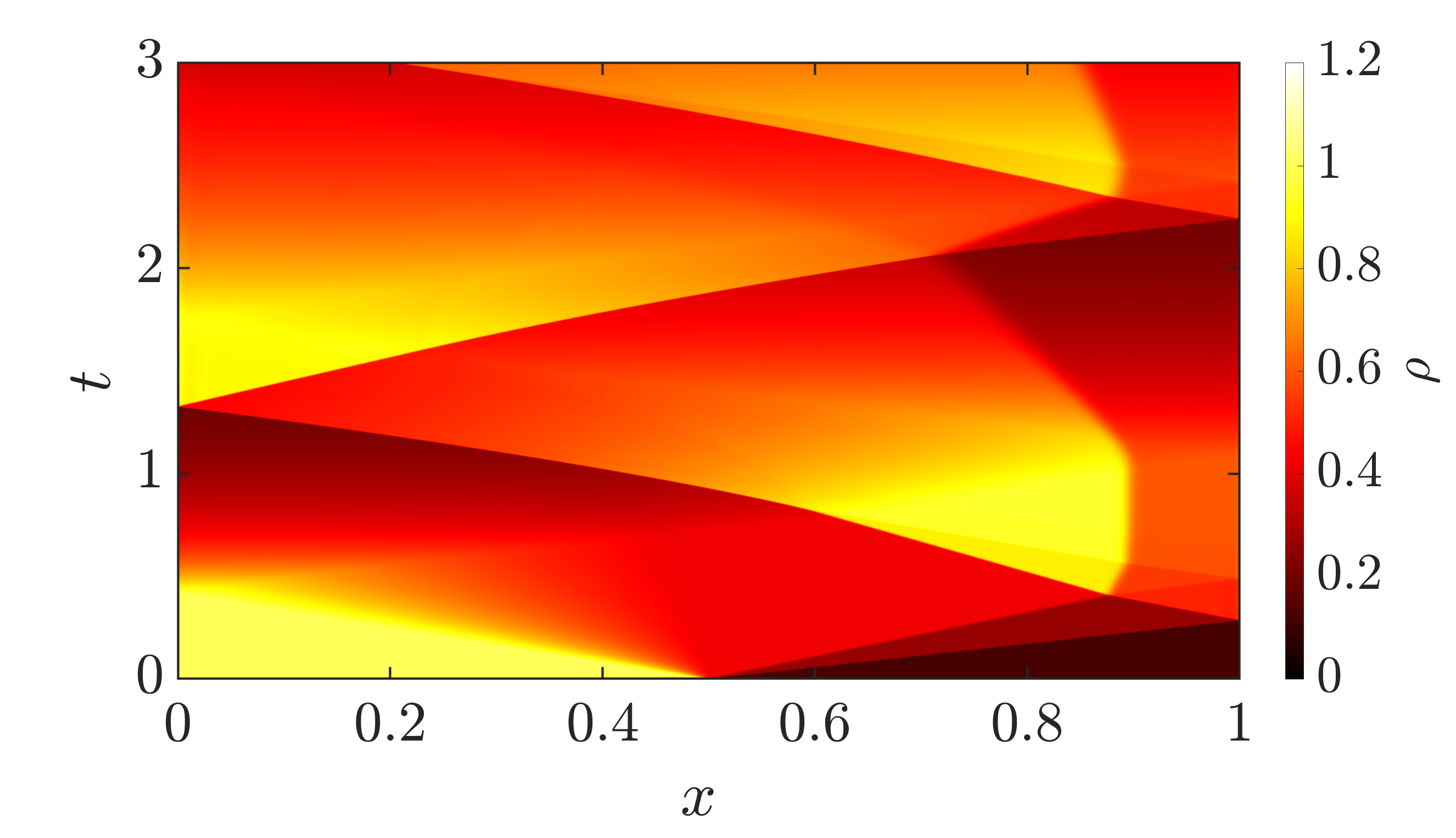}}
\subfloat[]{\includegraphics[width = 3.5in]{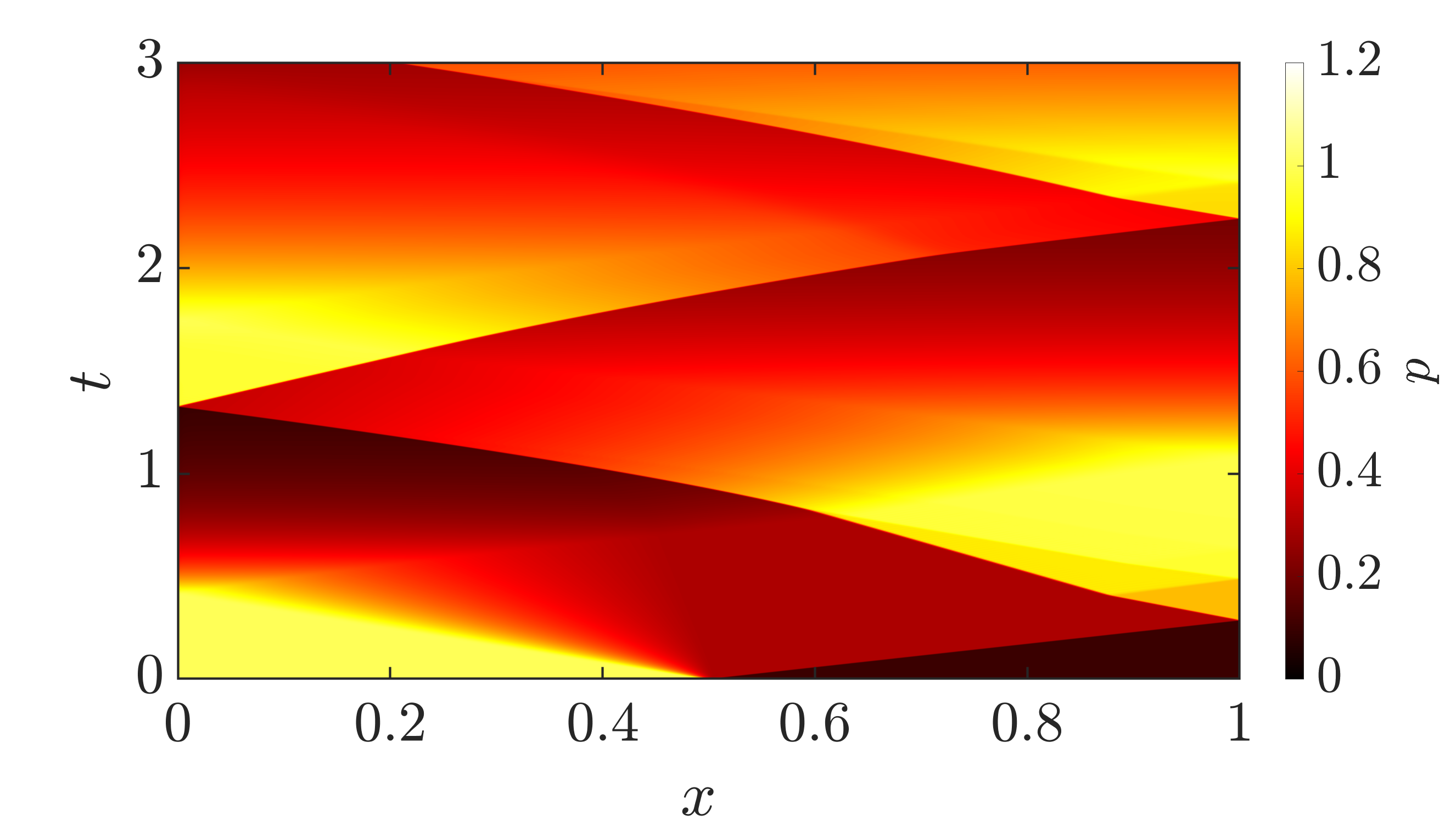}}
\caption{Density and pressure contours using Euler and MHD solvers on a stretched mesh with $600$ grid points and $\text{CFL}=0.5$ for $t=0$ to $3$. Wall boundary condition is applied at both ends. (a) and (b) show the density and pressure contours, respectively, using the Euler solver. (c) and (d) depict the density and pressure contours, respectively, using the MHD solver.}
\label{fig:TestCase1_2}
\end{figure}
\begin{figure}[!t]  
\hspace*{-1cm} 
\centering
\subfloat[]{\includegraphics[width = 3.0in]{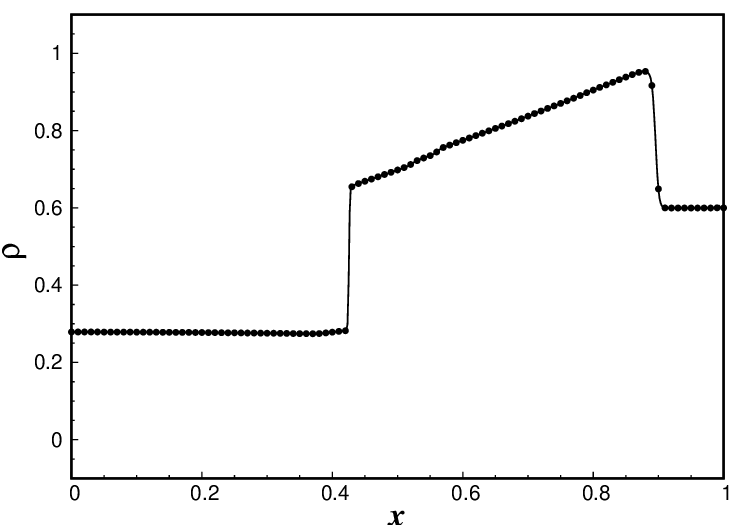}} \hspace*{0.5cm} 
\subfloat[]{\includegraphics[width = 3.0in]{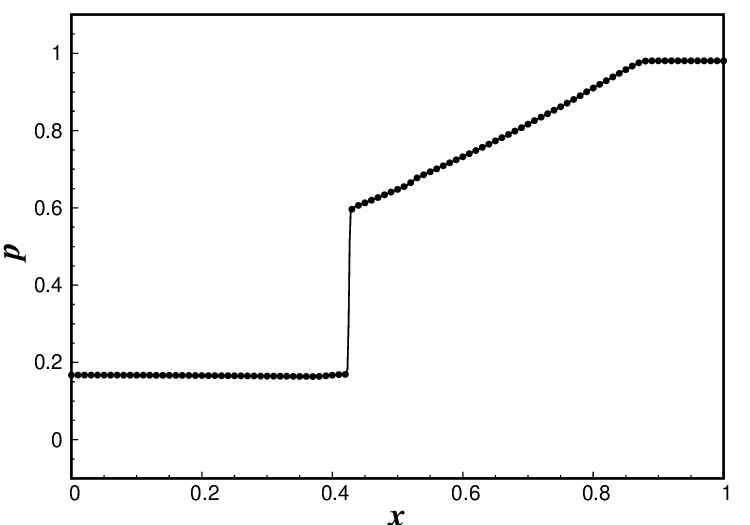}}\\
\caption{Density and pressure results using Euler and MHD solvers on a stretched mesh with $600$ grid points and $\text{CFL}=0.5$ at $t=0.9375$. (a) and (b) show the density and pressure results, respectively. \textemdash, using Euler solver; $\bullet$, using MHD solver. For the MHD solver, every 6th point of the results is shown.}
\label{fig:TestCase1_3}
\end{figure}

Figure \ref{fig:TestCase1_2} shows the time evolution of the density and pressure in the interval $t \in [0, 3]$. As can be observed, results from the Euler and MHD solvers are in excellent agreement. Moreover, propagating waves in the computational domain have the same patterns and the waves are correctly reflected from the wall. This observation suggests that the MHD characteristic boundary scheme works properly for the $\mathbf{B}=\textbf{0}$ case and the eigensystem is well defined for the non-magnetic case. 
 For a better comparison, the pressure and density fields obtained from the Euler and the MHD solvers at $t=0.9375$ \footnote{This time is chosen for the sake of a better presentation of the results.} are shown in the same plot in Fig. \ref{fig:TestCase1_3}. The results from the MHD solver are in close agreement with the Euler solver, which shows that the implemented MHD characteristic boundary scheme can correctly simulate the characteristic waves interaction with the wall boundary.

\subsection{Smooth Alfv\'en test with non-reflecting boundary condition}
\begin{figure}[!t]   
\centering
\hspace*{-2cm} 
\subfloat[]{\includegraphics[width = 4in]{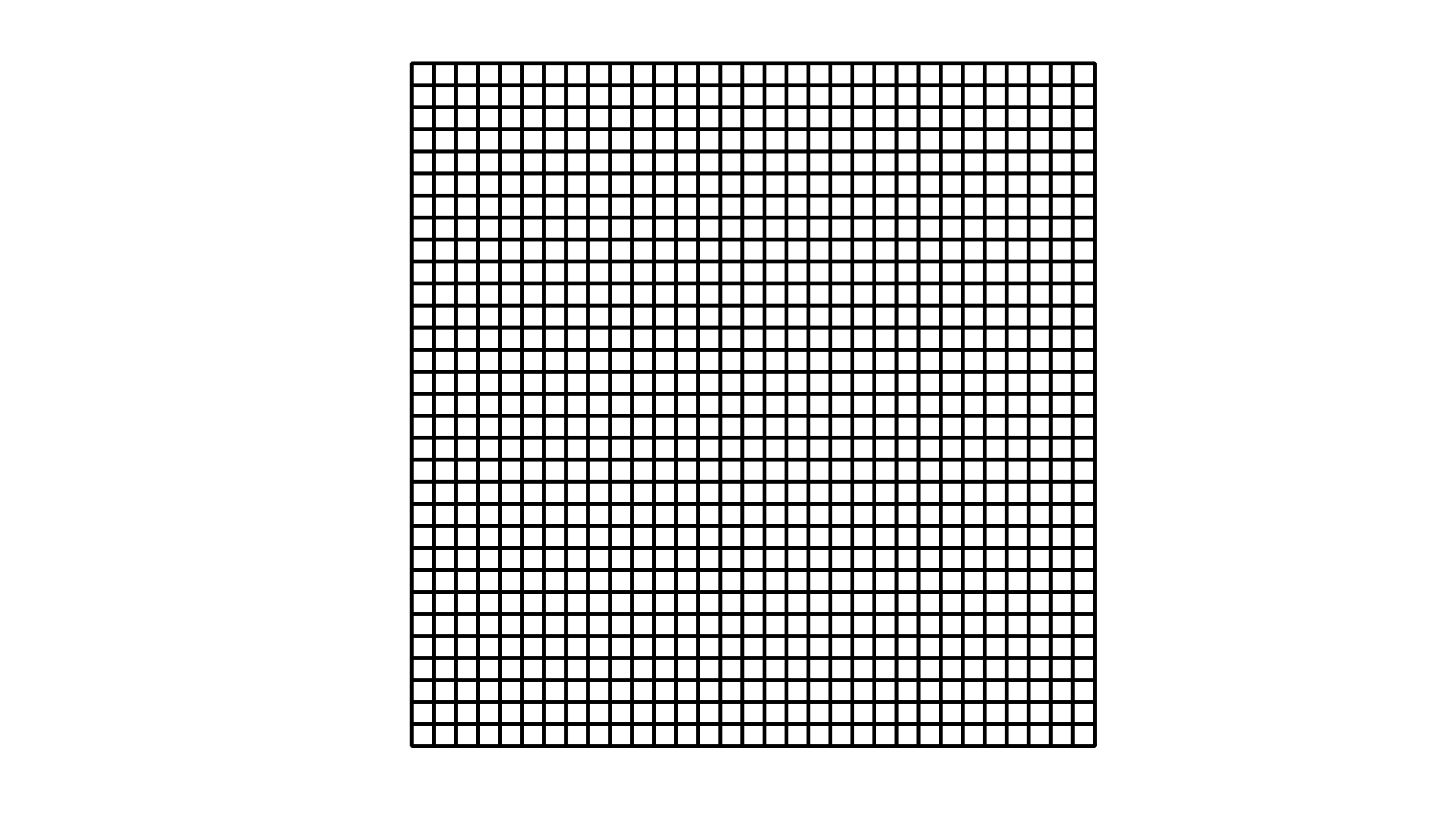}}\hspace*{-1cm} 
\subfloat[]{\includegraphics[width = 4in]{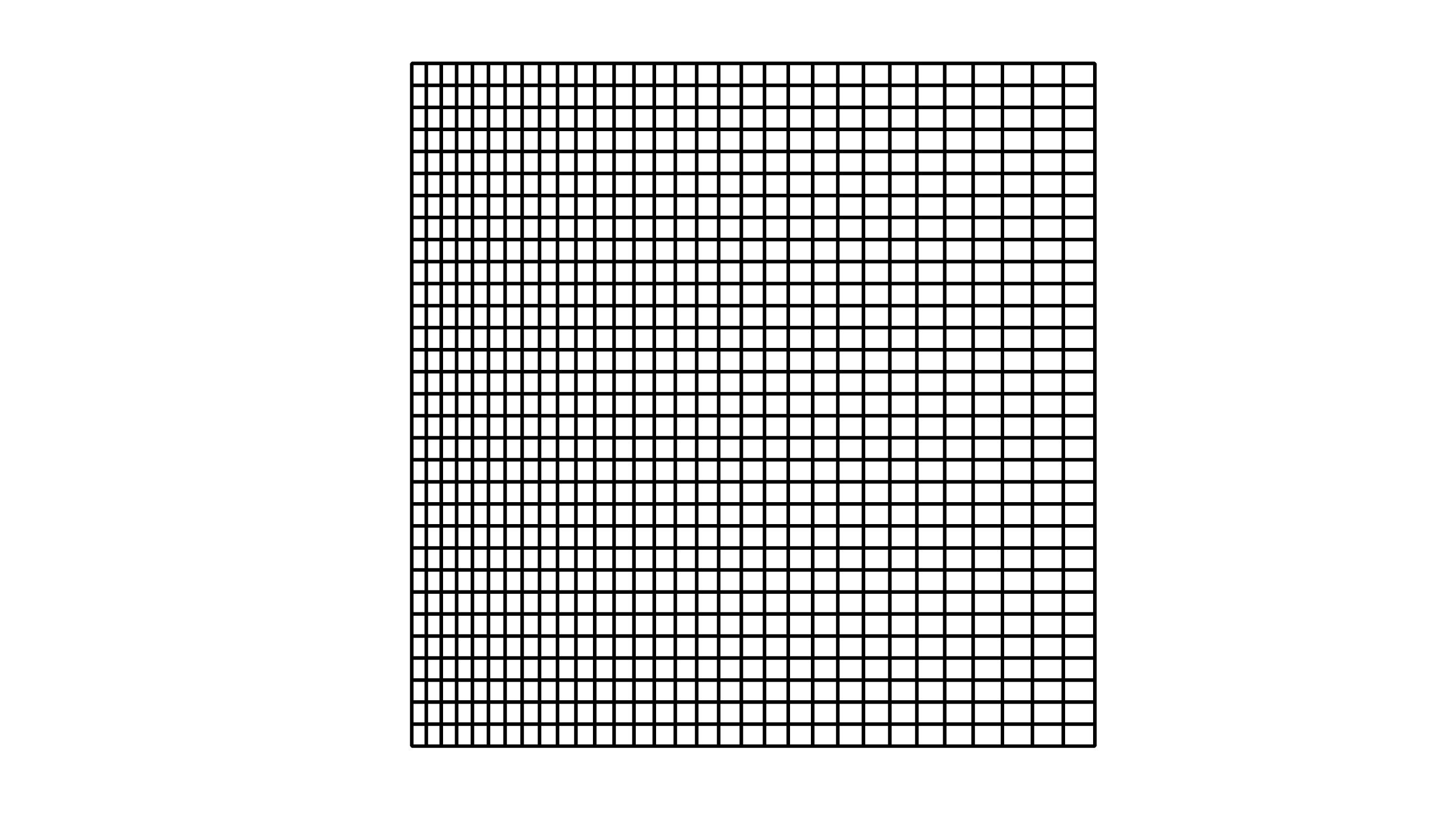}}\\
\centering
\subfloat[]{\includegraphics[width = 4in]{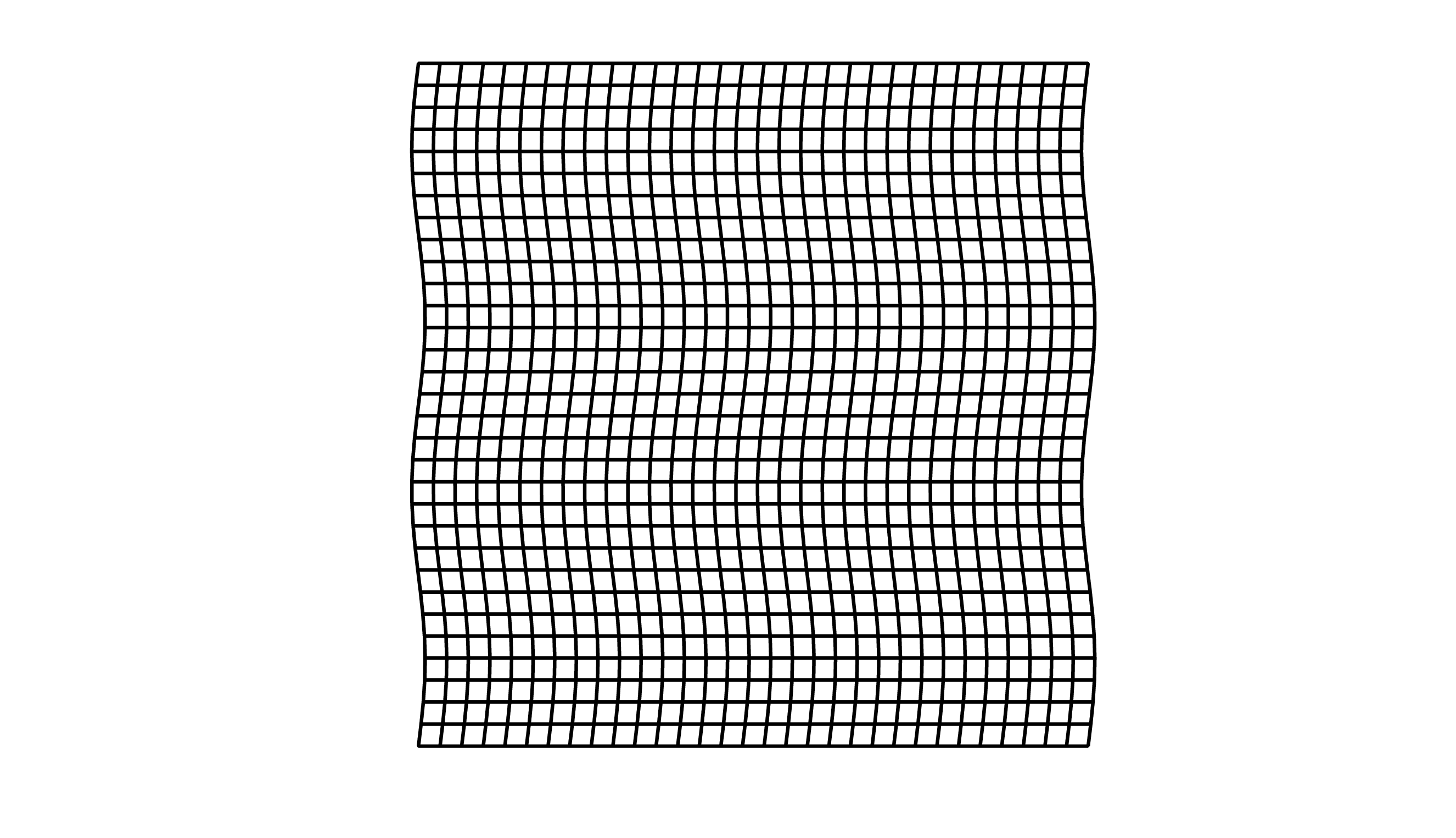}}
\caption{Three computational domains for solving the two-dimensional smooth Alfv\'en test with non-reflecting boundaries along $x$-direction. (a) shows uniform grid points in both $x$- and $y$-directions. Computational domain (b) has a non-uniform stretched grid points in $x$-direction using Eq. (\ref{eq:StretchedComputationalDomain}), and uniform grid points in $y$-direction. In computational domain (c), grid points along $x$-direction are perturbed according to Eq. (\ref{eq:CurvilinearComputationalDomain}) while $y$ grid points remain uniform. Every 4th node is shown.}
\label{fig:TestCase2_1}
\end{figure}

The smooth Alfv\'en test case is widely used in the literature to determine the order of accuracy of numerical schemes for the MHD problems with smooth flows. The initial condition is given as
\begin{equation}
      \left(\rho, u, v, w, p, B_x, B_y, B_z \right)=\left(1, 0, 0.1\sin\left(2 \pi x\right),  0.1, 0.1\cos\left(2\pi x\right), 1, 0.1\sin\left(2\pi x\right), 0.1\cos\left(2 \pi x\right)\right).
\end{equation}
In previous studies, periodic boundary conditions were applied along both $x$- and $y$-directions. Here, we employ the non-reflecting boundary condition along the $x$-direction and the periodic one along the $y$-direction. The analytical solution of the smooth Alfv\'en problem includes the Alfv\'en wave moving along the $x$-direction with a speed of $1$ and with constant density during the simulation. The computational and physical domains extend to $[0,1] \times [0,1]$; therefore by imposing a non-reflecting boundary condition along the $x$-direction after $t=1$, the Alfv\'en wave leaves the computational domain. As a result, the initial non-uniformities leave the domain and the magnetic field becomes uniform with $\mathbf{B}=\left(1, 0, 0.1\right)$. Hence, the final solution for $t \geq 1$ is given as 
\begin{equation}
     \left(\rho, u, v, w, p, B_x, B_y, B_z \right)=\left(1, 0, 0, 0.1, 0.1, 1, 0, 0.1 \right).
\end{equation}
In this test case, three computational domains with $128 \times 128$ grid size are selected to investigate the efficiency of the non-reflecting characteristic boundary scheme. The first computational domain includes uniform grid points in both $x$- and $y$-directions, whereas the second one consists of non-uniform stretched grid points along the $x$-direction using Eq. (\ref{eq:StretchedComputationalDomain}) and uniform grid points in the $y$-direction. The third computational domain is set to be $[\xi, \eta]^2 \in [0,1] \times [0, 1]$ with the perturbed $x$-grid lines according to the following mapping function \citep{Christlieb&Feng1999}
\begin{equation}\label{eq:CurvilinearComputationalDomain}
    x= \xi + \alpha_x \sin \left(2 \pi \, \eta \, \beta_x \right),
\end{equation}
where $\alpha_x$ shows the magnitude of the perturbation and $\beta_x$ is the wave number of the perturbation \citep{Christlieb&Feng1999} with uniform $y$-grid points. In this study, parameters $\alpha_x$ and $\beta_x$ are defined the same as the ones mentioned by Christlieb et al. \citep{Christlieb&Feng1999}, where $\alpha_x$ and $\beta_x$ are equal to $0.01$ and $2$, respectively. These three computational domains are shown in Fig. \ref{fig:TestCase2_1}.\par
As explained in section \ref{sec:nonrefBC}, a non-reflecting boundary condition is imposed by setting the amplitude of the incoming waves to zero \citep{Thompson1987}. In this test case, we investigate the implementation and accuracy of the non-reflecting boundary condition using the MHD characteristic boundary scheme. The sixth-order compact and fourth-order Runge-Kutta schemes are used for the spatial discretization and time marching, respectively, with CFL$=0.6$. 
\par
The calculated Root Mean Square (RMS) of the error for the magnetic field components is less than $0.05\%$ for the uniform computational domain. The RMS of the errors is calculated using
\begin{equation}
    \text{RMSE}=\sqrt{\frac{\Sigma_{i=1} ^N \left(f_{\mathrm{anal}_i}-f_{\mathrm{num}_i} \right)^2}{N}},
\end{equation}
where $f_{\mathrm{anal}_i}$ and $f_{\mathrm{num}_i}$ show the analytical and numerical results at grid point $i$, respectively, and $N$ is the total number of grid points. 
For the stretched grid, the results perfectly followed the analytical ones and the error is almost $0.1\%$ for the magnetic field components. Small discrepancies are observed compared to the analytical solution for the curvilinear computational domain, since the characteristic boundary scheme is derived while assuming the flow to be locally one-dimensional. This assumption is not valid for the curvilinear mesh and thus, these errors are inevitable. The calculated RMSE in the curvilinear domain for $B_x$, $B_y$, $B_z$, and $p$ are $0.23\%$, $0.21\%$, $0.22\%$, and $0.27\%$, respectively. Calculated errors for the three computational domains are reported in Table (\ref{table:TestCase2_1}).
Furthermore, it is also beneficial to calculate the $L^{\infty}$ error for the curvilinear computational domain since it shows the maximum error and evaluate the scheme's performance on the grid points with the greatest distortion. The obtained $L^{\infty}$ error for $B_x$, $B_y$, $B_z$, and $p$ on the curvilinear domain are 0.44\%, 0.47\%, 0.33\%, and 0.29\%, respectively. As expected, the calculated $L^{\infty}$ errors are higher than $L^2$ ones. However, since the computed errors are under 0.5\%, we can conclude that the scheme works well on boundary grids with the greatest misalignment as well.
\begin{table*}
\centering
\caption{\label{table:TestCase2_1} Calculated RMSE for the two-dimensional smooth Alfv\'en wave test case with the non-reflecting boundaries along $x$-direction, using three computational domains.}
\begin{tabular}{ccccc}
\\
\hline
Computational domain & Error in $p$ & Error in $B_x$ & Error in $B_y$ & Error in $B_z$\\
\hline
(a)  & $\leq 0.06 \%$ &  $\leq 0.05 \%$ & $\leq 0.04 \%$ & $\leq 0.05 \%$\\
 (b) & $0.13 \%$ & $0.1 \%$ & $0.11 \%$ & $0.1 \%$\\
(c) & $0.27 \%$ & $0.23 \%$ & $0.21 \%$ & $0.22 \%$\\
 \hline
\end{tabular}
\end{table*}

\par
\begin{figure}[!t]  
\centering
\hspace*{-1cm} 
\subfloat[]{\includegraphics[width = 3in]{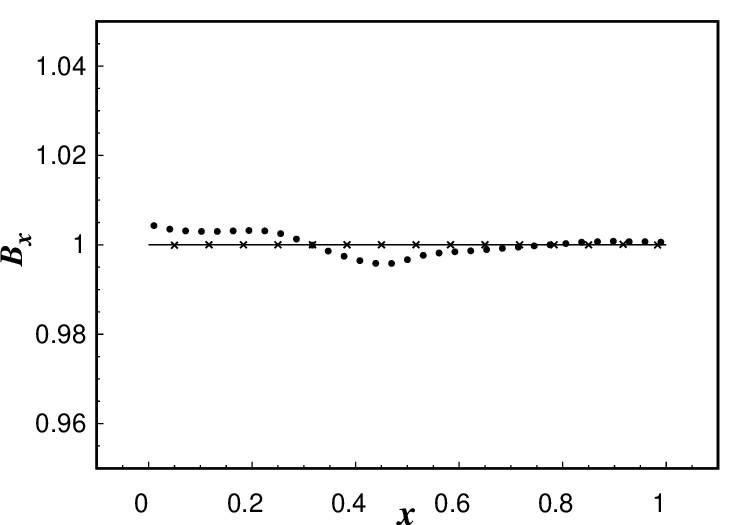}} \hspace*{0.5cm} 
\subfloat[]{\includegraphics[width = 3in]{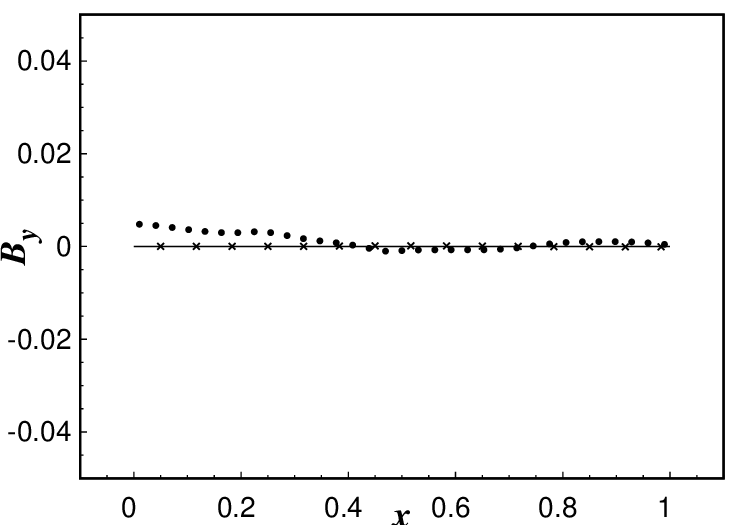}}\\
\hspace*{-1cm}
\subfloat[]{\includegraphics[width = 2.9in]{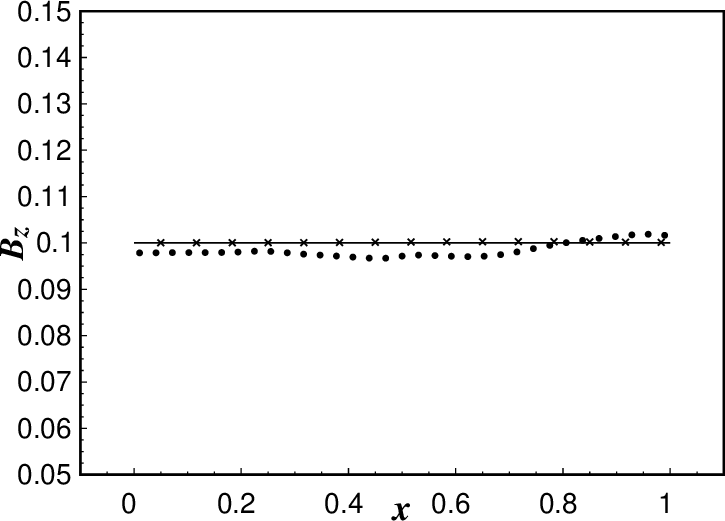}} \hspace*{0.5cm}
\subfloat[]{\includegraphics[width = 2.9in]{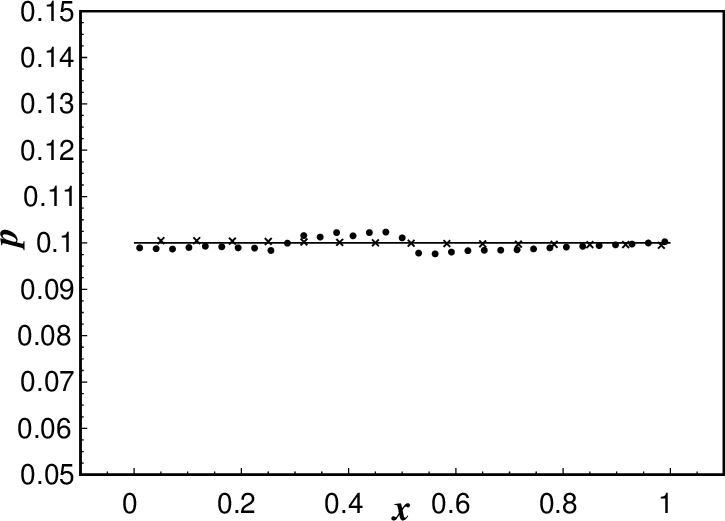}}
\caption{Comparing the magnetic field and pressure results using three computational domains for the two-dimensional smooth Alfv\'en problem with non-reflecting boundaries along $x$ and periodic boundary along $y$, at $t=1$ and $y=0.5$. \textemdash, uniform computational domain; $\times$, non-uniform stretched computational domain; $\bullet$, curvilinear computational domain. For the stretch and the curvilinear grids, every 5th and 4th point of the results is shown, respectively.}
\label{fig:TestCase2_2}
\end{figure}

Figure \ref{fig:TestCase2_2} shows the comparison of the magnetic field and pressure results using three computational domains. Results from the uniform and non-uniform computational domains are indistinguishable on the same plot, while for the curvilinear case, small discrepancies are observed.

\subsection{Gaussian pulse test with wall boundary condition}
\begin{figure}[!t]  
\centering
\hspace*{-1cm} 
\subfloat[]{\includegraphics[width = 3in]{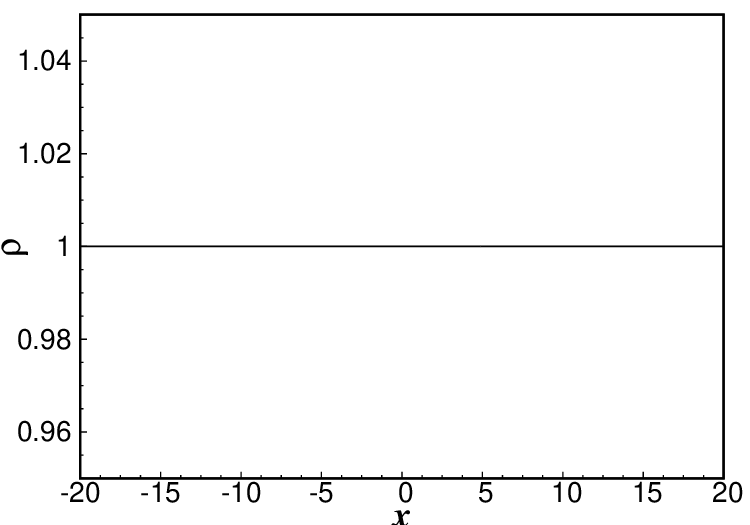}} \hspace*{0.5cm} 
\subfloat[]{\includegraphics[width = 3in]{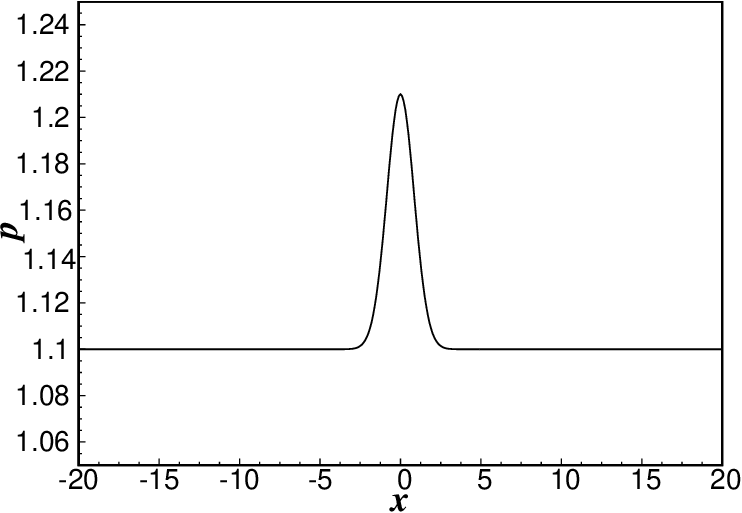}}\\
\hspace*{-1.0cm} 
\subfloat[]{\includegraphics[width = 3in]{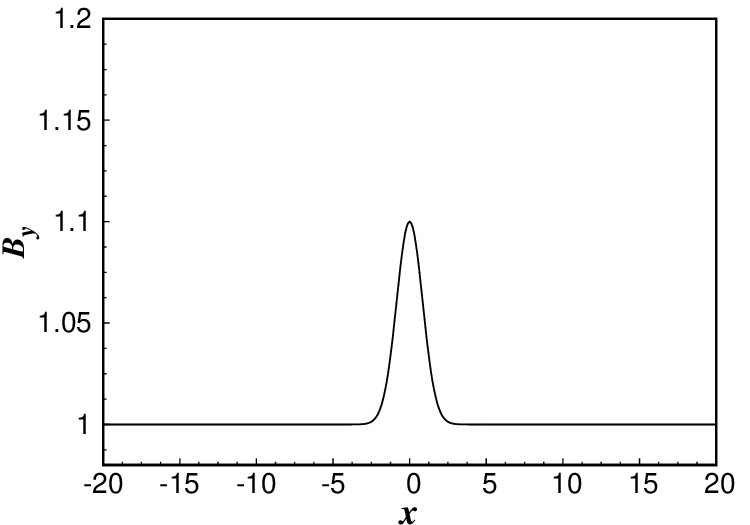}} \hspace*{0.5cm} 
\subfloat[]{\includegraphics[width = 3in]{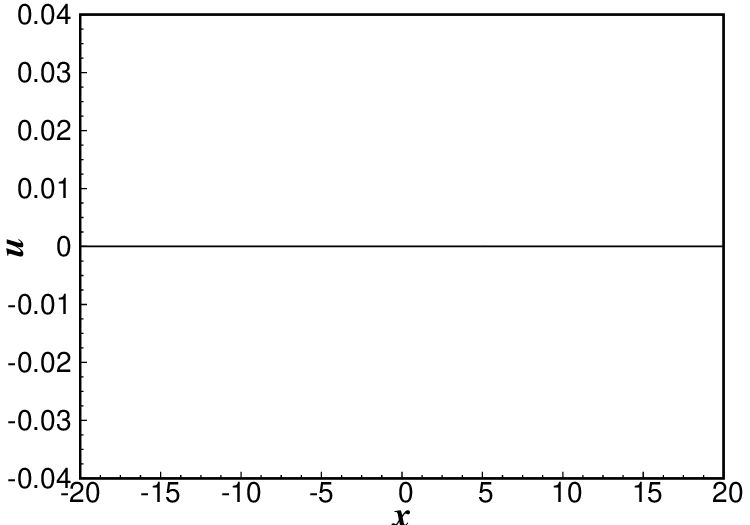}}
\caption{The one-dimensional Gaussian pulse problem using $400$ grid points with wall boundaries, at $t=0$. (a) density, (b) pressure, (c) transverse component of the magnetic field, and (d) velocity.}
\label{fig:TestCase3_1}
\end{figure}

This one-dimensional test case is designed to investigate the wall boundary condition for the MHD problem using the implemented characteristic boundary scheme. The initial condition reads as
\begin{equation}
(\rho, u, v, B_x, B_y, p)=\left(1.0, 0, 0, 0.75, 1+\epsilon \exp\left[\frac{-x^2}{b^2}\right],1.1\left(1+\epsilon \exp\left[\frac{-x^2}{b^2}\right]\right)\right),
\end{equation}
where $\epsilon=0.1$, and $b=1.2$. As clearly seen in Fig. \ref{fig:TestCase3_1}, the initial condition consists of a Gaussian pulse perturbation applied to the pressure and the transverse magnetic field at $x=0$. The no-penetration wall boundary condition is applied at both ends.
The normal velocity is forced to zero at the wall. According to Eq. (\ref{eq:MHDchar2}), we need to set $\mathcal{L}_6$ equal to $\mathcal{L}_3$ and $\mathcal{L}_8$ to $\mathcal{L}_1$ at the left boundary and the other way around for the right boundary to satisfy the wall boundary condition. This procedure ensures that $\partial u/\partial t$ remains zero during the simulation, which is essential for this type of boundary condition. 
\par
The simulation is conducted in the non-uniform stretched grid according to Eq. (\ref{eq:StretchedComputationalDomain}) mapping function, with $400$ grid points in the computational domain, $\xi \in [-20, 20]$, and $\text{CFL}=0.6$. The sixth-order compact scheme and the fourth-order Runge-Kutta scheme are used for the spatial discretization and time marching, respectively. Moreover, this test case is simulated in the uniform computational domain to compare the results with its non-uniform counterpart. 
\par
\begin{figure}[!t]  
\centering
\hspace*{-1cm} 
\subfloat[]{\includegraphics[width = 3in]{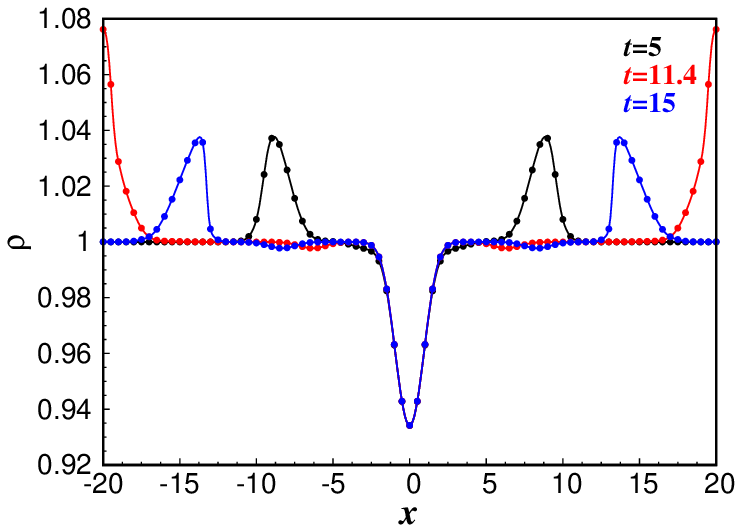}} \hspace*{0.5cm}
\subfloat[]{\includegraphics[width = 3in]{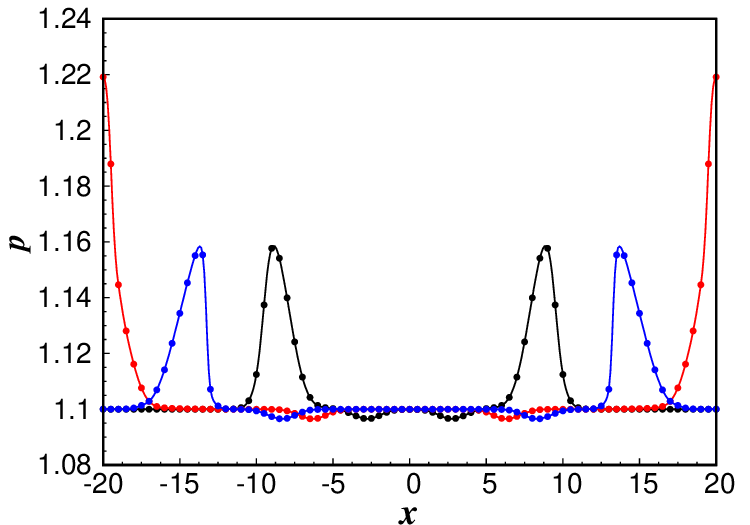}}\\
\hspace*{-1.1cm} 
\subfloat[]{\includegraphics[width = 3in]{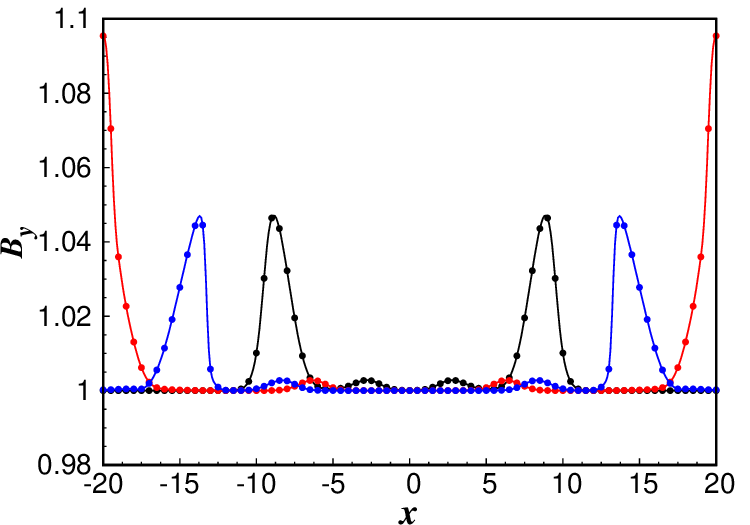}} \hspace*{0.5cm}
\subfloat[]{\includegraphics[width = 3in]{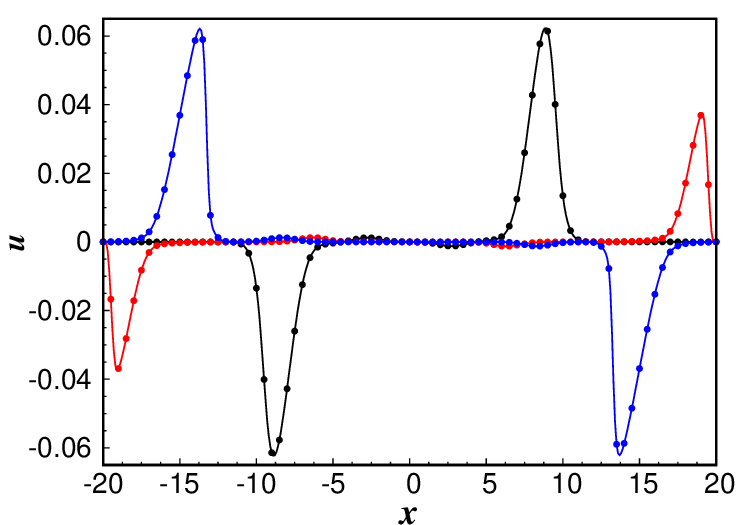}}
\caption{The one-dimensional Gaussian pulse problem with wall boundaries; \textemdash, non-uniform computational domain; $\bullet$, uniform computational domain with $400$ grid points at $t=5$ before waves interact with the walls (in black), $t=11.4$ as the waves interact with the walls (in red), and $t=15$ after the waves reflected from the walls (in blue). Every 5th point of the results is shown for the uniform case. (a) density, (b) pressure, (c) transverse component of the magnetic field, and (d) velocity.}
\label{fig:TestCase3_2}
\end{figure}

Figure (\ref{fig:TestCase3_2}) show the evolution of the density, pressure, transverse magnetic, and velocity fields before and after the interaction of the waves with the walls. As can be observed, there is a good agreement between the uniform and non-uniform results. This suggests that the characteristic boundary scheme works properly, independent of the grid configuration.
\par
Since in the initial condition, the flow speed is assumed to be zero, the results are symmetric with respect to the origin, $x=0$, for the pressure, density, and the transverse magnetic fields. For the velocity field, the left- and right-going characteristic waves are observed having the same speed, albeit with different signs.
\par
Figure \ref{fig:TestCase3_2} displays the results at $t=5$ in black, before the interaction of the waves with the wall boundaries. At $t=5$, the propagation of the left and right, fast and slow magnetosonic waves can be observed. There is a pressure bump at the origin at $t=0$ and, hence, the density value should decrease at $x=0$ to compensate the pressure disturbance at this point. 
Figure \ref{fig:TestCase3_2} shows the results as the left and right going waves interact with the wall at $t=11.4$ in red. The magnitude of the pressure increases at both ends due to the presence of the wall. 
Figure \ref{fig:TestCase3_2} depicts the reflected waves from the wall boundaries at $t=15$ in blue. The reflected waves are symmetric with respect to the origin, as expected. Moreover, the reflected density and pressure peaks have the same values as those before their interaction with the wall. These observations validate the implemented wall boundary condition.

\subsection{Magnetosonic wave propagation test}
\subsubsection{One-dimensional case}
The one-dimensional magnetosonic wave propagation test case is designed to investigate the single wave inlet boundary condition. This test case demonstrates the procedure of using the characteristic boundary scheme to inject a single magnetosonic wave into the domain at a boundary. Subsequently, the evolution of the injected characteristic wave into the computational domain can be studied. The initial condition, whose uniformity helps us to better verify the time evolution of the injected characteristic wave, is given as
\begin{equation}\label{eq:TestCase4_1}
(\rho, u, v, B_x, B_y, p)=\left(1.0, 0, 0, 0.75, 1.0, 0.5\right),
\end{equation}
with the single fast magnetosonic wave inlet and non-reflecting outlet boundary conditions applied to the left and right boundaries, respectively. For this test case, the specific heat value is assumed to be $\gamma=1.4$.
\par
In order to have a single fast characteristic wave entering the domain, we should set the magnitude of all the incoming waves to zero, except the fast one. Here, we assume a Gaussian pulse is moving towards the left boundary with the fast magnetosonic speed. To implement the fast magnetosonic inlet boundary condition, we defined the $\mathcal{L}_8$ operator, the characteristic operator with respect to the fast characteristic wave, as:
\begin{equation}\label{eq:TestCase4_2}
    \mathcal{L}_8=-\epsilon \exp\left[\frac{- \left(c_{\mathrm{f}} \, t-1 \right)^2}{2b^2}\right],
\end{equation}
where variables $\epsilon$ and $b$ are assumed to be 0.5 and 0.2, respectively. Furthermore, $c_{\mathrm{f}}$ shows the velocity of the injected fast wave. Since $\mathcal{L}_8$ is the only incoming wave, the time evolution of the pressure can be written as
\begin{equation}\label{eq:TestCase4_3}
    \frac{\partial p}{\partial t}+ \alpha_{\mathrm{f}} \, \rho \, a^2 \mathcal{L}_8=0.
\end{equation}
Therefore, evolution of the pressure is given by the error function, i.e., the integral of the Gaussian pulse function. Similarly, the time evolution of the other primitive variables are
\begin{equation}\label{eq:TestCase4_4}
     \frac{\partial \rho}{\partial t}+ \alpha_{\mathrm{f}} \, \rho \mathcal{L}_8=0,
\end{equation}
\begin{equation}\label{eq:TestCase4_5}
    \frac{\partial u}{\partial t}+ \alpha_{\mathrm{f}} \, c_{\mathrm{f}} \mathcal{L}_8=0,
\end{equation}
and,
\begin{equation}\label{eq:TestCase4_6}
    \frac{\partial B_y}{\partial t}+ \alpha_{\mathrm{s}} \, c_{\mathrm{s}}\,\sqrt{\rho} \, \beta_y \, a \mathcal{L}_8=0.
\end{equation}
According to Eqs. (\ref{eq:TestCase4_4}) to (\ref{eq:TestCase4_6}), we expect the results for the density, velocity, and the transverse magnetic fields to be in the form of the error function as well. 
\par
This test is also repeated for the case of injecting a Gaussian pulse with the slow magnetosonic speed into the computational domain at the left boundary, with the same initial condition. Analogous to the previous case, we again set the magnitude of all the incoming waves to zero, with the exception of the slow one. Hence, the $\mathcal{L}_6$ operator is defined as
\begin{equation}\label{eq:TestCase4_7}
    \mathcal{L}_6=-\epsilon \exp\left[\frac{- \left(c_{\mathrm{s}} \, t-1 \right)^2}{2b^2}\right],
\end{equation}
where variables $\epsilon$ and $b$ are assumed to be 0.5 and 0.2, respectively.
\par
\begin{figure}[!t]   
\centering
\hspace*{-1cm} 
\subfloat[]{\includegraphics[width = 3in]{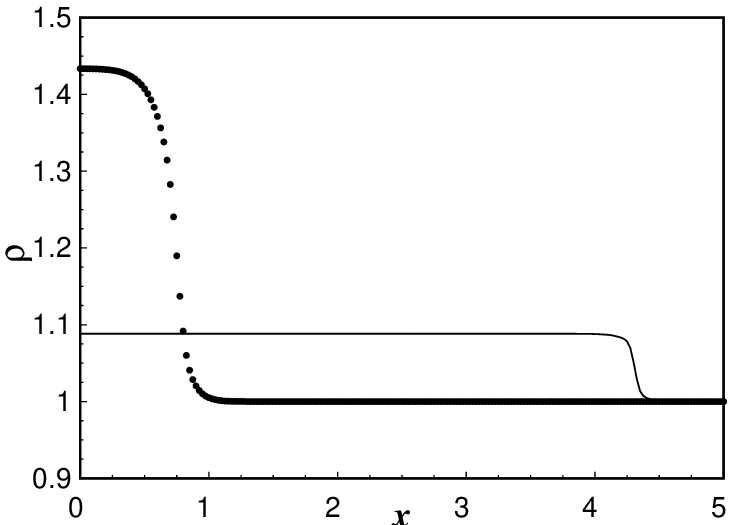}} \hspace*{0.5cm} 
\subfloat[]{\includegraphics[width = 3in]{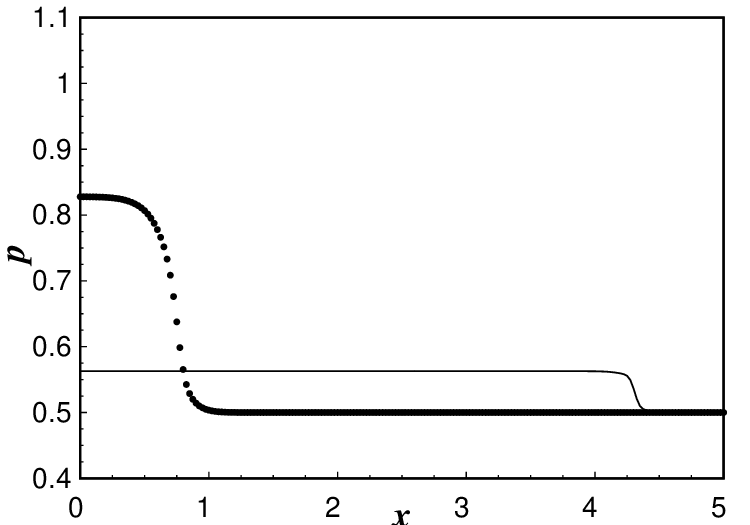}}\\
\hspace*{-1cm}
\subfloat[]{\includegraphics[width = 3in]{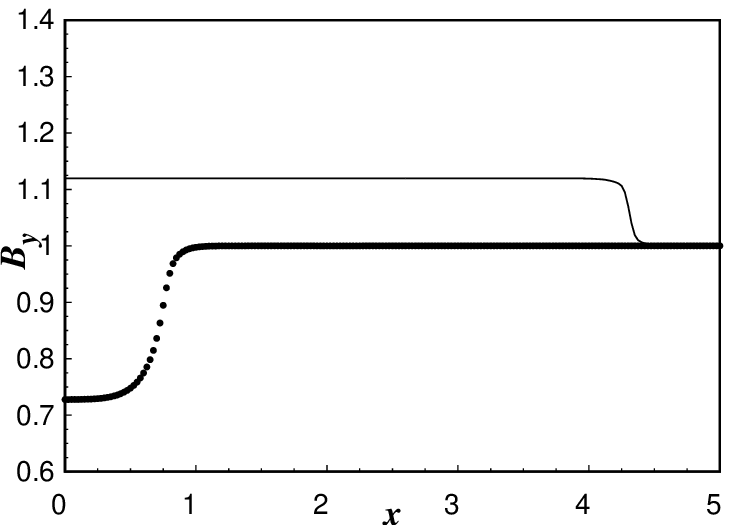}} \hspace*{0.5cm}
\subfloat[]{\includegraphics[width = 3in]{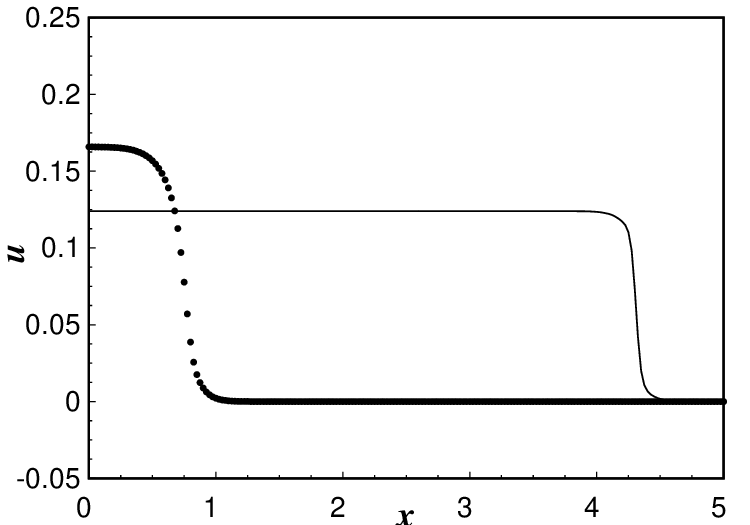}}
\caption{The magnetosonic waves propagation problem on a non-uniform stretched grid with $200$ points at $t=3.5$. (a) density, (b) pressure, (c) transverse component of the magnetic field, and (d) velocity. \textemdash, fast magnetosonic wave propagation; $\bullet$, slow magnetosonic wave propagation.}
\label{fig:TestCase4_1}
\end{figure}

The simulation of both the slow and fast characteristic wave injections are conducted on a computational domain with the non-uniform stretched grid points constructed according to Eq. (\ref{eq:StretchedComputationalDomain}), with $200$ grid points and $\text{CFL}=0.5$, for $x \in [0, 5]$. Results for these two cases are shown in Fig. \ref{fig:TestCase4_1}.

\par
As expected, for the fast case, the injected disturbance travels faster in the domain, and therefore the initial density, velocity, pressure, and magnetic fields evolve more rapidly. The calculated mean speed value for the slow and fast magnetosonic waves are $0.435$ and $1.439$, respectively. Hence, at $t=3.5$, we expect the position lag between the onset of the error functions to be approximately $3.5$, which can be observed in Fig. \ref{fig:TestCase4_1} as well.
\par
The difference exhibited in the amplitudes of the results are due to the difference in the dependence of the time derivatives of the primitive variables on the amplitudes of the waves. Furthermore, as already seen in Eq. (\ref{eq:MHDchar6}), the coefficients of the $\mathcal{L}_6$ and $\mathcal{L}_8$ operators have different signs. Therefore, the signs of the induced error functions for the transverse magnetic fields are opposite.
\par
The accuracy of the results can also be verified by studying the density and magnetic fields reaction to the perturbation injected into the domain. As mentioned earlier, the density and magnetic fields are in phase for the fast magnetosonic wave. This behaviour can be observed in Fig. \ref{fig:TestCase4_1} as they both have the same trend. However, for the slow magnetosonic wave, density and magnetic fields are out of phase. This is why their behaviour is opposite for the slow magnetosonic test case.
\par
\begin{figure}[!t]
\centering
\hspace*{-1cm} 
\subfloat[]{\includegraphics[width = 3in]{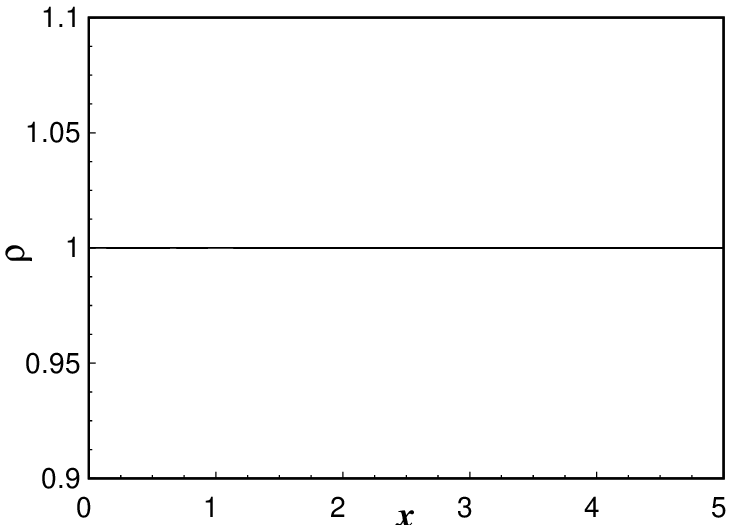}} \hspace*{0.5cm}
\subfloat[]{\includegraphics[width = 3in]{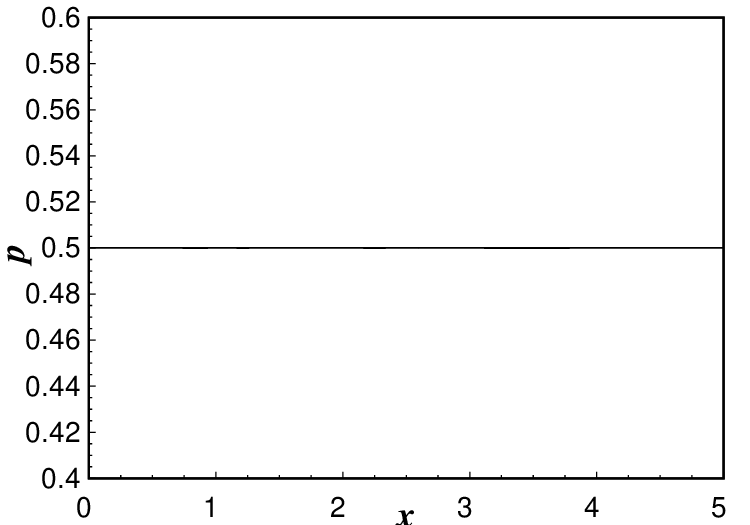}}\\
\hspace*{-1cm}
\subfloat[]{\includegraphics[width = 3in]{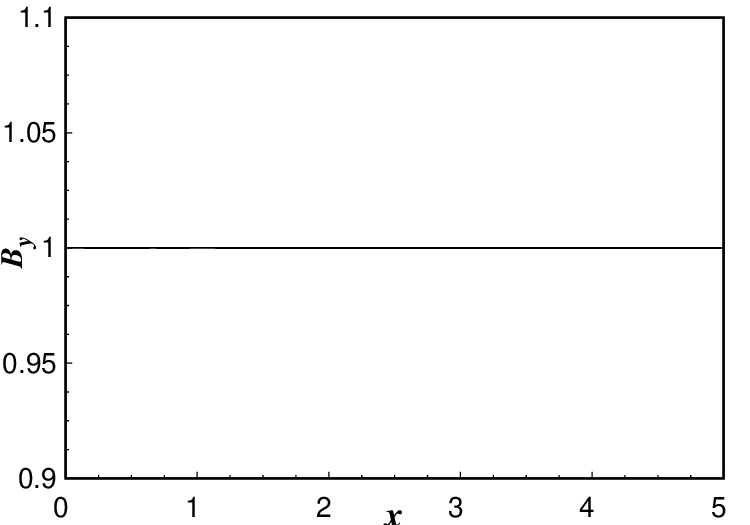}} \hspace*{0.5cm}
\subfloat[]{\includegraphics[width = 3in]{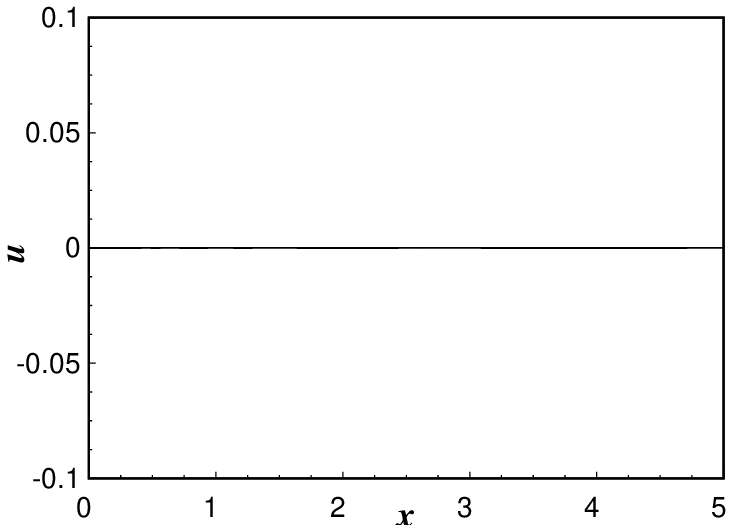}}
\caption{The Alfv\'en wave propagation problem on a non-uniform stretched grid with $200$ points at $t=1.0$. (a) density, (b) pressure, (c) transverse component of the magnetic field, and (d) velocity.}
\label{fig:TestCase4_2}
\end{figure}

We also studied this test case by injecting an Alfv\'en wave at the left boundary. As the Alfv\'en wave is incompressible, we expect its effect on density to be null. In order to apply this boundary condition, we define the $\mathcal{L}_7$ operator as
\begin{equation}
    \mathcal{L}_7=-\epsilon \exp\left[\frac{- \left(c_{\mathrm{a}} \, t-1 \right)^2}{2b^2}\right],
\end{equation}
where variables $\epsilon$ and $b$ are set similarly as the previous test. 
Figure \ref{fig:TestCase4_2} shows the results of this test case. In the time evolution relations for the density, velocity, and pressure, Eqs. (\ref{eq:MHDchar1}), (\ref{eq:MHDchar2}), and (\ref{eq:MHDchar8}), the $\mathcal{L}_7$ operator does not appear, and, hence, has no effect on these variables. On the other hand, the time evolution of the transverse magnetic field can be recast as
\begin{equation}
    \frac{\partial B_y}{\partial t}-\sqrt{\rho} \, \beta_z \, S \left(\mathcal{L}_2 + \mathcal{L}_7 \right)=0.
\end{equation}
Therefore, the $\mathcal{L}_7$ operator affects the transverse magnetic field but since $B_z=0$, the coefficient of the Alfv\'en characteristic waves becomes zero. Consequently, as shown in Fig. \ref{fig:TestCase4_2}, the transverse magnetic field does not change at all.

\subsubsection{Two-dimensional case}
The fast magnetosonic wave propagation test case is also conducted on a two-dimensional computational domain similar to Fig. \ref{fig:TestCase2_1}c with $[x, y] \in [0, 5] \times [0, 5]$ and the same initial condition as Eq. (\ref{eq:TestCase4_1}). At the left boundary along the $x$-direction, a fast magnetosonic wave is injected into the domain, while the right boundary is set to be a non-reflecting outlet. Periodic boundaries are applied along the $y$-direction. The simulation is performed on $200 \times 100$ grid points and the CFL number is $0.6$.

\par
\begin{figure}[!t]
\centering
\hspace*{-1.5cm} 
\subfloat[]{\includegraphics[width = 3in]{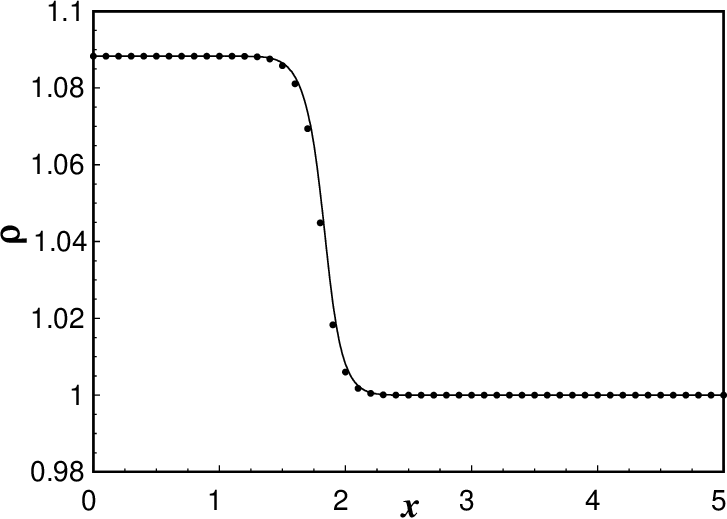}} \hspace*{0.5cm}
\subfloat[]{\includegraphics[width = 3in]{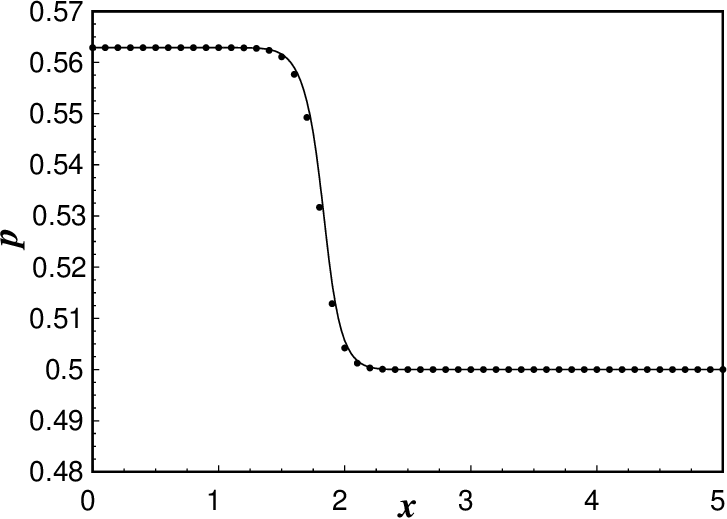}}\\
\hspace*{-1.5cm}
\subfloat[]{\includegraphics[width = 3in]{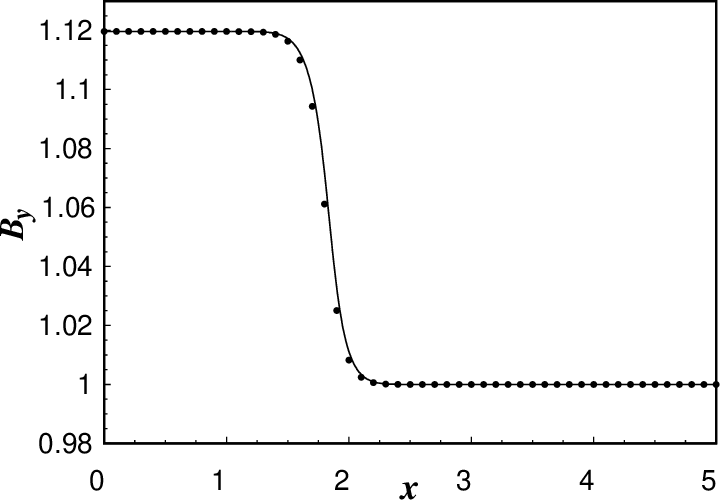}} \hspace*{0.5cm} 
\subfloat[]{\includegraphics[width = 3in]{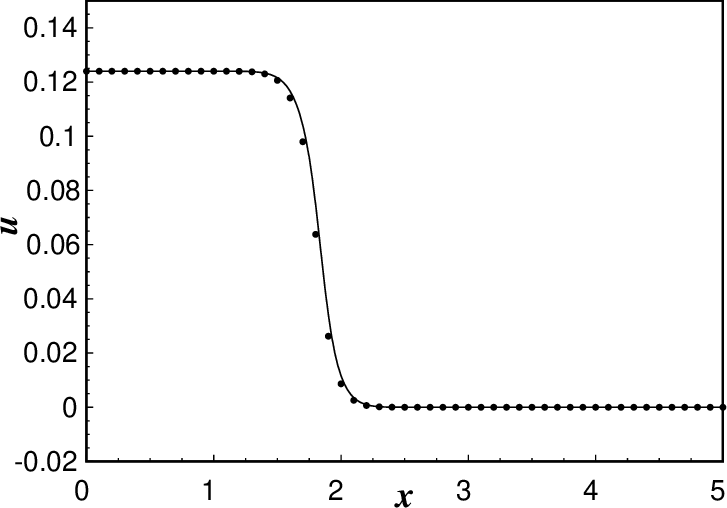}}
\caption{The two-dimensional fast magnetosonic wave propagation problem using $200 \times 100$ grid points on a perturbed computational domain at $t=1.86$. (a) density, (b) pressure, (c) transverse component of the magnetic field, and (d) velocity. \textemdash, results from two-dimensional test case at $y=2.5$; $\bullet$, results from one-dimensional case. For the one-dimensional test case, every 4th point is shown.}
\label{fig:TestCase4_3}
\end{figure}

 Results from the perturbed two-dimensional computational domain at $y=2.5$ are compared with those from the one-dimensional case, shown in Fig. \ref{fig:TestCase4_3} at $t =1.86$
\footnote{This time is chosen for the sake of a better presentation of the results.}. The calculated RMSE for the density, pressure, velocity, and transverse magnetic fields from the two-dimensional simulation relative to the one-dimensional case are $0.30\%$, $0.21\%$, $0.42\%$, and $0.41\%$, respectively. Therefore, we conclude that the single wave injection boundary condition works properly on the curvilinear domain as well. The primary sources of the captured errors for this test case are the perturbed boundary points along the $x$-direction, which are functions of $\xi$ and $\eta$. This computational domain is not consistent with the locally one-dimensional assumption that we made while implementing the characteristic boundary scheme. Therefore, these errors are prone to occur.

\subsection{Brio--Wu shock tube test with non-reflecting boundary condition}

This test case was first proposed by Brio and Wu \citep{Brio&Wu1987}. The initial condition is given as:
\begin{equation}
      \left(\rho, u, v, w, p, B_x, B_y, B_z \right)= \begin{cases}
            (1, 0, 0, 0, 1, 0.75, +1, 0) \    \  \  \   \  \  \ \  \  \ \  \ \text{\        \ for $x < 0$}, \\ 
            (0.125, 0, 0, 0, 0.1, 0.75, -1, 0) \         \ \    \  \text{for $x > 0$},  
            \end{cases}
\end{equation}
with the heat capacity ratio $\gamma=2$. The Brio--Wu problem is a standard test for MHD codes since it accurately represents the shocks, rarefactions, contact discontinuities, and compound structure of the MHD problem. This test case exhibits strong shocks and discontinuities in the presence of a magnetic field, and thus comprises a rigorous test case for the proposed characteristic boundary scheme for MHD problems that are of interest to applications such as fusion energy.
\par
Due to the existence of sharp discontinuities and shocks, it is necessary to apply Total Variation Diminishing (TVD) schemes for temporal integration and spatial discretization. That being the case, in this problem, the third-order TVD Runge-Kutta \citep{Gottlieb&Shu1998} and fifth-order WENO methods have been utilized for the time integral and spatial discretization, respectively. Furthermore, non-reflecting boundaries are applied at the right and left boundaries for this case. The non-reflecting boundary condition will help us assess whether the implemented characteristic boundary method can efficiently distinguish the incoming and outgoing waves at the inlet and outlet, especially when shocks and jump discontinuities reach the boundaries. Furthermore, the validity of the derived eigenvectors and eigenvalues can be evaluated for problems with discontinuities. The test case is simulated on the computational domain consisting of $800$ non-uniform stretched grid points according to Eq. (\ref{eq:StretchedComputationalDomain}) with $x \in [-1,1]$. The CFL number is set to $0.6$ with the simulation time $t=0$ to $t=3$. \par
\begin{figure}[t!]
\hspace{-2cm}
\includegraphics[width = 1.2\textwidth]{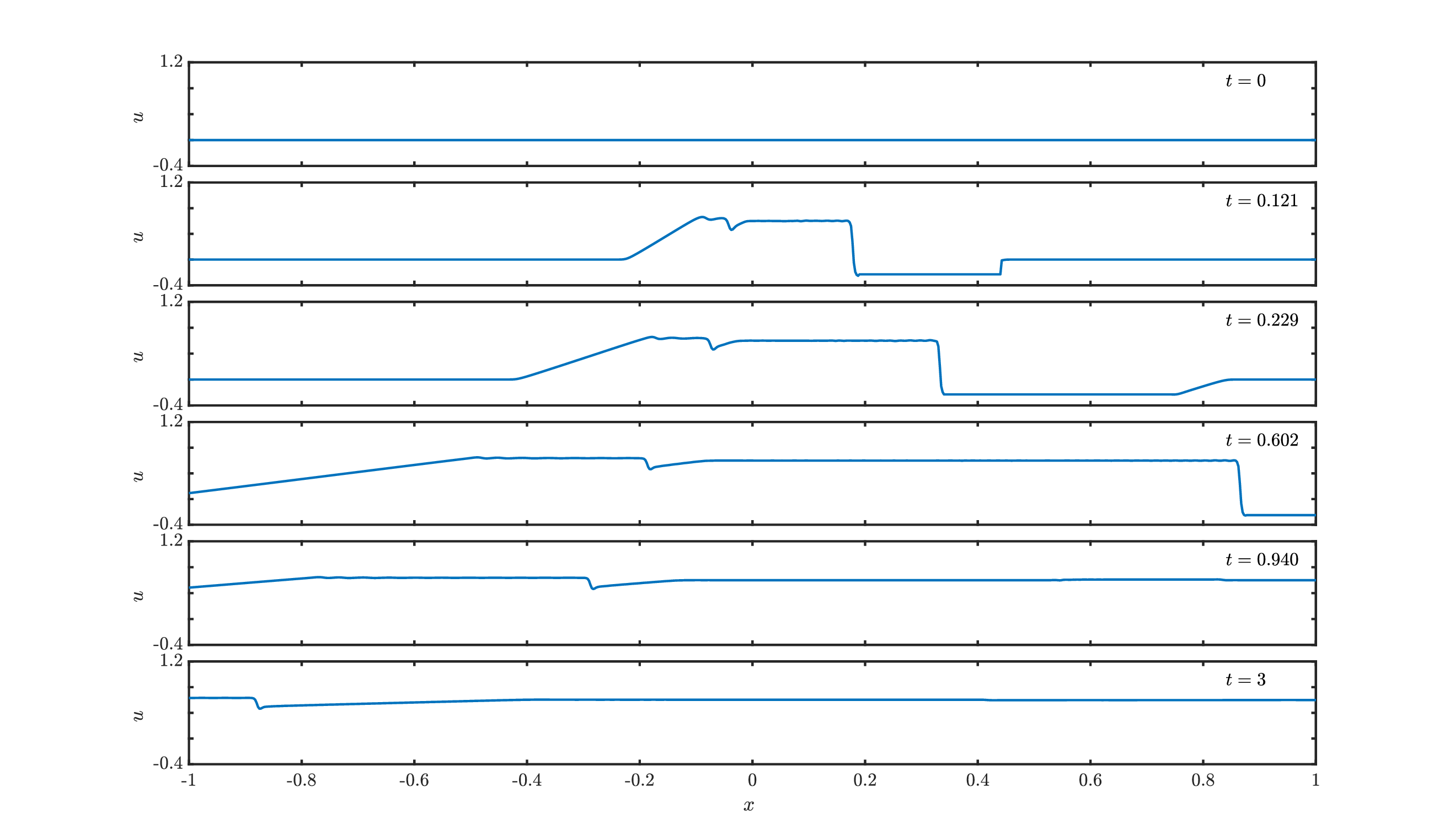}
\caption{The velocity field of the Brio--Wu shock tube test case with non-reflecting boundaries on a stretched mesh with $800$ grid points and $\text{CFL}=0.6$ at six different time steps, $t=0$, $0.121$, $0.229$, $0.602$, $0.940$, and $3$.}
\label{fig:TestCase5_1}
\end{figure}
\begin{figure}[t!]
\hspace{-2cm}
\includegraphics[width = 1.2\textwidth]{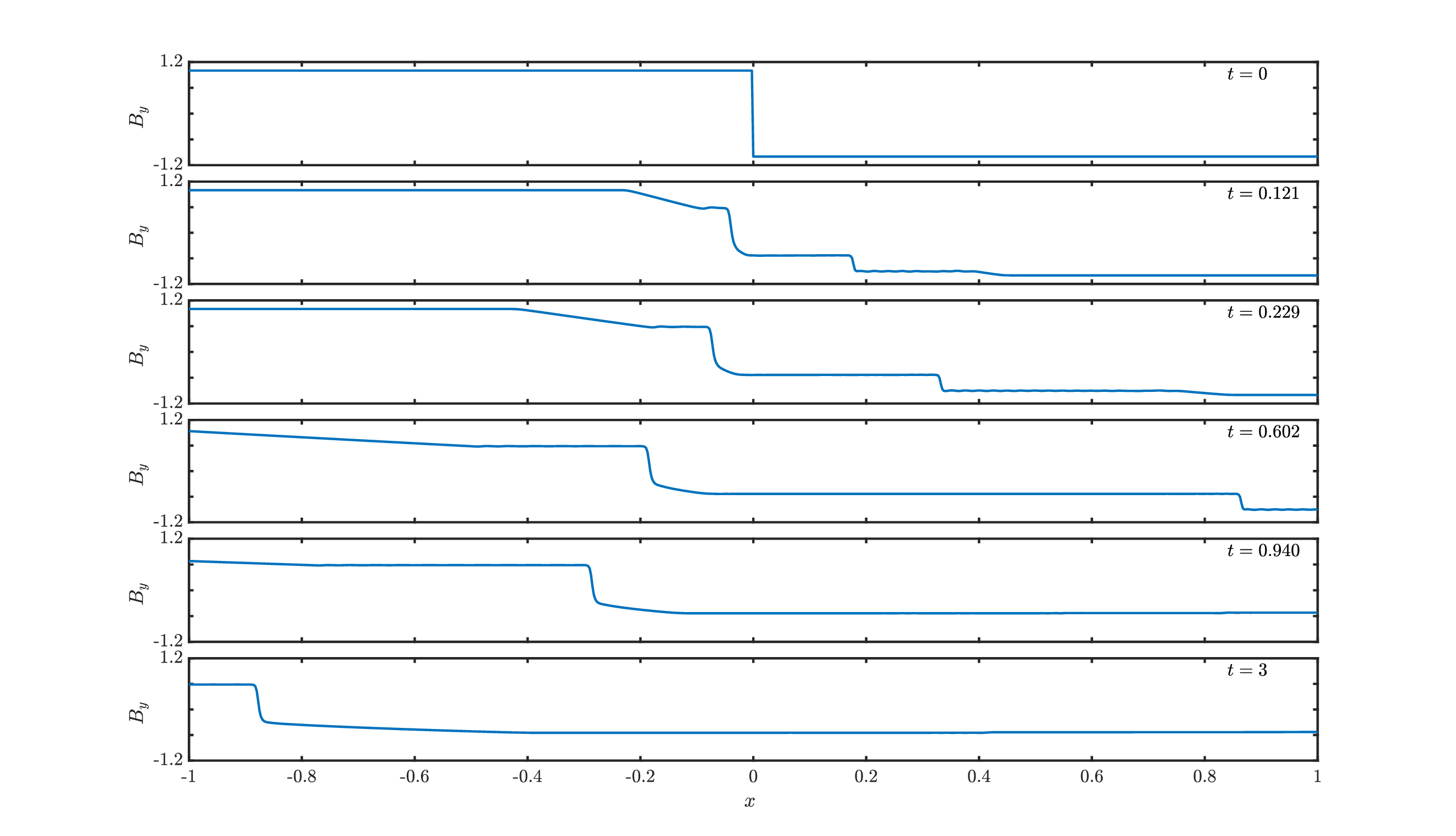}
\caption{The transverse magnetic field of the Brio--Wu shock tube test case with non-reflecting boundaries on a stretched mesh with $800$ grid points and $\text{CFL}=0.6$ at six different time steps, $t=0$, $0.121$, $0.229$, $0.602$, $0.940$, and $3$.}
\label{fig:TestCase5_2}
\end{figure}

The solution of the Brio--Wu shock tube consists of a right-traveling magneto-fast rarefaction wave, which is followed by a magneto-slow shock wave and a contact discontinuity. Moreover, two waves, i.e., magneto-fast rarefaction and magneto-slow compound wave, propagate to the left. The compound wave contains an intermediate shock followed by a slow rarefaction wave. Figure (\ref{fig:TestCase5_1}) depicts the evolution of the velocity field at six different time steps, $t=0$, $0.121$, $0.229$, $0.602$, $0.940$, and $3$. At $t=0.121$, both right- and left-traveling shock and rarefaction waves can be observed. At $t=0.229$, the fast rarefaction wave propagating to the right gets closer to the boundary. According to Fig. (\ref{fig:TestCase5_1}), at $t=0.602$, the right-traveling rarefaction wave has already left the right boundary without any reflection. Moreover, the other rarefaction wave moving towards the left boundary is also leaving the domain. At $t=0.940$, the right-traveling slow shock wave has completely left the computational domain and the left-traveling rarefaction wave is also on the verge of crossing the left boundary entirely. At the last time step, $t=3$, the left-traveling compound wave has gotten closer to the left boundary, as we expected, and the left-traveling rarefaction wave has completely exited the domain. Figure (\ref{fig:TestCase5_2}) represents the evolution of the transverse magnetic field at the same time steps. Similarly, it can be observed that the right- and left-traveling rarefactions and shocks leave the domain at the left and right non-reflecting boundaries without causing any numerical instability due to the interaction of discontinuities with the boundaries. Furthermore, it can be concluded that the proposed characteristic boundary scheme performs robustly in the presence of shocks as well, and the derived eigenstructure remains valid.

\section{Conclusion}
Imposing proper boundary conditions is of crucial importance for the accurate simulation of plasma dynamics. In this study, we introduce characteristic boundary conditions for the numerical solution of the MHD equations on multi-dimensional, curvilinear grids. A parallel, high-order code, with sixth-order accuracy in space and fourth-order accuracy in time, was developed for the simulation of the compressible MHD equations. A fifth-order WENO scheme was also implemented to capture discontinuities such as shock waves. 
\par
The correctness of the derived scheme was first verified for the Sod's shock tube problem, a non-magnetic test case.
Then, a set of test cases were designed to investigate the accuracy of the proposed scheme, in particular non-reflecting boundary conditions which are of crucial importance for the accurate simulations of instabilities. We have also shown how to inject single-mode waves into the computational domain. Such a boundary condition can be used to independently study the dynamics of each wave mode.
In conclusion, the studied test cases showed the promise and correctness of the implemented MHD characteristic boundary scheme for one- and two-dimensional problems in both cartesian and curvilinear coordinate systems.

\section*{Acknowledgments}
The authors thank Andrew Higgins for his invaluable suggestions and comments to improve this manuscript. This study was supported by the Natural Science and Engineering Research Council of Canada (NSERC) through an NSERC Discovery Grant.

\section*{Appendix A. Convergence study}
\section*{Two-Dimensional Smooth Alfv\'en Test}
\begin{figure}[]   
\centering
\includegraphics[width = 5in]{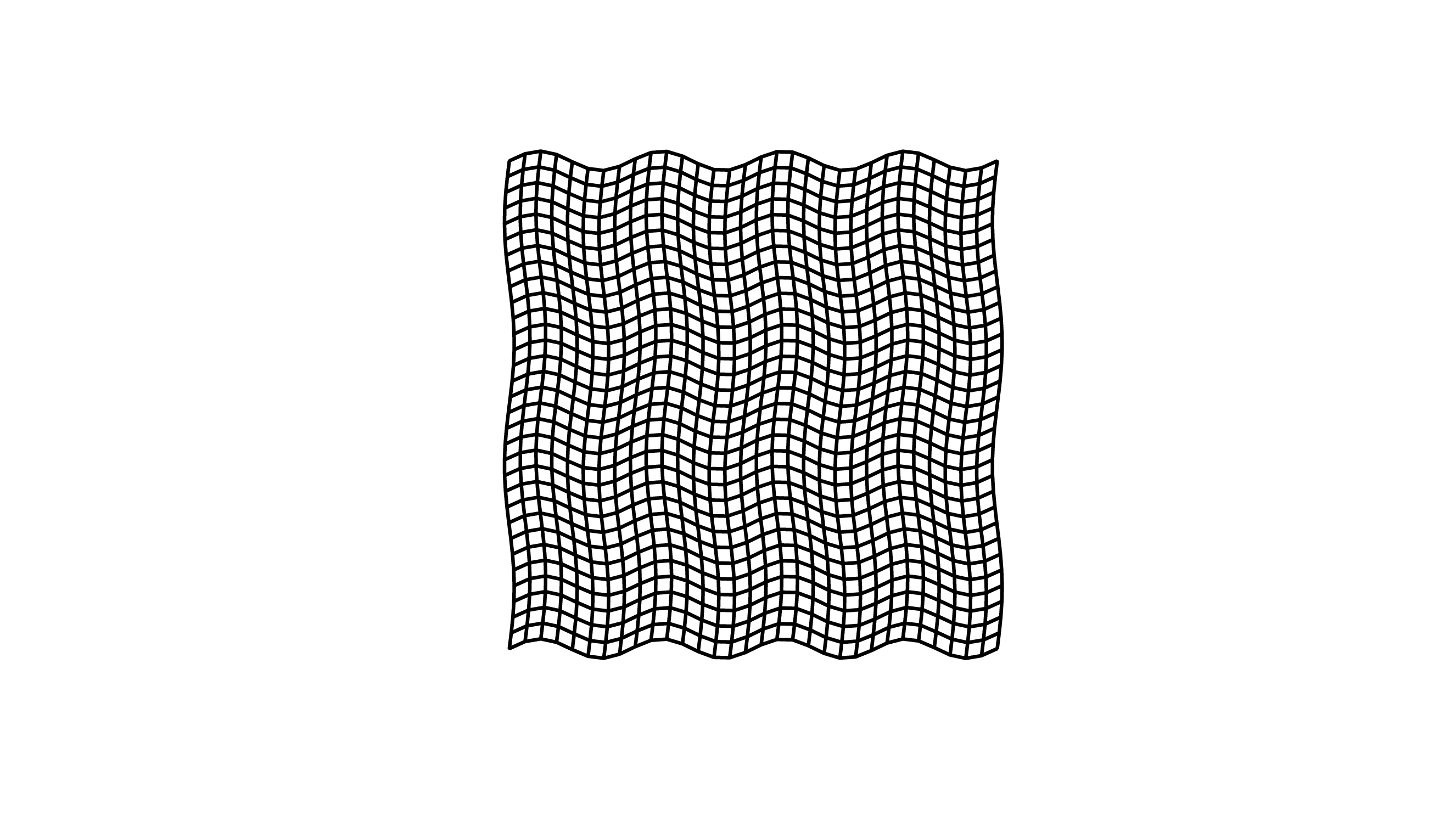}
\vspace{-1.0cm}
\caption{Computational domain for the two-dimensional smooth Alfv\'en wave problem with periodic boundary conditions along the $x$- and $y$-directions. The grid size of $32 \times 32$ is shown with $[\xi, \eta]^2 \in [0,1] \times [0, 1]$.}
\label{fig:perturbed}
\end{figure}

This problem is widely used to verify the accuracy order of the numerical scheme for MHD problems with smooth flows. The capability of Alfv\'en waves to move over long times and distances in numerical solvers is important especially in turbulence simulations 
\cite{Derigs&Winters2016}. Alfv\'en waves can be damped as a result of numerical dissipation in the scheme. If the code fails to capture these waves, the turbulence behaviour cannot be correctly studied, as MHD turbulence is mainly sustained by Alfv\'en waves\cite{Balsara2014}. 
The initial condition is given as:
\begin{equation}
     \left(\rho, u, v, w, p, B_x, B_y, B_z \right)=\left(1, 0, 0.1\sin\left(2 \pi x\right),  0.1, 0.1\cos\left(2\pi x\right), 1, 0.1\sin\left(2\pi x\right), 0.1\cos\left(2 \pi x\right)\right),
\end{equation}
with the periodic boundary condition along the $x$- and $y$-directions. The computational domain is set to be $[\xi, \eta]^2 \in [0,1] \times [0, 1]$ with the perturbed grid lines according to the mapping functions \citep{Christlieb&Feng1999}
\begin{equation}
    x= \xi + \alpha_x \sin\left(2 \pi \,\eta \, \beta_x \right),
\end{equation}
and
\begin{equation}
    y= \eta + \alpha_y \sin\left(2 \pi \, \xi \, \beta_y\right),
\end{equation}
where $\alpha_x$ and $\alpha_y$ show the magnitude of the perturbation and $\beta_x$ and $\beta_y$ are the wave numbers of the perturbation \citep{Christlieb&Feng1999}. In this study, these parameters are taken to be the same as the ones mentioned by Christlieb \textit{et al.} \citep{Christlieb&Feng1999}, where $\alpha_x$, $\alpha_y$, $\beta_x$, and $\beta_y$ are equal to $0.01$, $0.02$, $2$, and $4$, respectively. The computational domain is shown in Fig. (\ref{fig:perturbed}).
\par
The exact solution is an Alfv\'en wave propagating along the $x$-direction with a speed of 1. Therefore, the initial circularly polarized Alfv\'en wave propagates across a periodic domain. The numerical results, using both spatial discretization schemes, the central compact and the WENO schemes, are compared to the analytical solution at $t=5$. Different grid resolutions with the constant $\text{CFL}=0.5$ were used for calculating the order of accuracy, as reported in Tables (\ref{Tab:A1}) and (\ref{Tab:A2}). 
In Tables (\ref{Tab:A1}) and (\ref{Tab:A2}), the $L^{\infty}$ error values are shown as well as the estimated convergence rates. 
For each resolution, the order of accuracy was calculated with respect to the previous courser one.
As can be concluded from Tables \ref{Tab:A1} and \ref{Tab:A2}, fifth-order WENO and sixth-order compact schemes are almost fourth-order and fifth-order globally accurate, respectively.
\begin{table}
\centering
  \caption{$L^{\infty}$ errors for two-dimensional smooth Alfv\'en wave problem using the fifth-order WENO scheme.}
\label{Tab:A1}
  \begin{tabular}[c]{ |p{3cm}|p{3cm}|p{3cm}|  }
  \hline
 \multicolumn{3}{|c|}{WENO scheme } \\
 \hline
 Mesh & Error in $\mathbf{B}$ & Order\\
 \hline
 \hline
 32 $\times$ 32 & 5.4128-3 & \textbf{------}\\
 \hline
 64 $\times$ 64 & 4.3418e-4 & 3.64\\
 \hline
 128 $\times$ 128 & 2.8093e-5 & 3.95\\
 \hline
 256 $\times$ 256 & 1.7680e-6 & 3.99\\
 \hline
\end{tabular}
\end{table}

\begin{table}
  \begin{center}
   \caption{$L^{\infty}$ errors for two-dimensional smooth Alfv\'en wave problem using the sixth-order compact scheme.}
 \label{Tab:A2}
  \begin{tabular}{ |p{3cm}|p{3cm}|p{3cm}|  }
 \hline
 \multicolumn{3}{|c|}{Compact scheme} \\
 \hline
 Mesh & Error in $\mathbf{B}$ & Order\\
 \hline
 \hline
  32 $\times$ 32 & 4.9879e-3 & \textbf{------}\\
 \hline
 64 $\times$ 64 & 1.9298e-4 & 4.66\\
 \hline
 128 $\times$ 128 & 6.2869e-6 & 4.94\\
 \hline
 256 $\times$ 256 & 1.9920e-7 & 4.98\\
 \hline
\end{tabular}
  \end{center}
\end{table}

\section*{Appendix B. Simplifying MHD eigenstructure}

As was highlighted before, three types of waves termed fast, intermediate (Alfv\'en), and slow are formed according to their characteristic speeds in the MHD system. Depending on the magnitude and direction of the magnetic field, these waves may coincide with each other. Therefore, the MHD equations may involve multiplicities of eigenvalues, provided the modes coincide. For example, this phenomenon may occur in the following conditions for the one-dimensional case \citep{Brio&Wu1988,Brio&Wu1987}:\\
\indent 1) When $B_x=0$ and $c_{\mathrm{s}}=c_{\mathrm{a}}=0$. Therefore, the eigenvalue $u$ becomes of  multiplicity five.\\
\indent 2) When ${B_y}^2+{B_z}^2=0$, ${c_{\mathrm{s}}}^2=\min(a^2, {b_x}^2)$, and ${c_{\mathrm{f}}}^2=\max(a^2, b_x^2)$. Thus for the case $a^2 \neq {b_x}^2$, either if ${c_{\mathrm{s}}}^2={b_x}^2$ or ${c_{\mathrm{f}}}^2={b_x}^2$, the multiplicity of $u\pm c_{\mathrm{a}}$ becomes two. Also, for the case ${c_{\mathrm{s}}}^2={c_{\mathrm{f}}}^2={b_x}^2$, the multiplicity of $u \pm c_{\mathrm{a}}$ changes to three. \\
Consequently, the MHD governing equations do not form the strictly hyperbolic system of equations. Owing to this property, we should make sure that the eigensystem of the MHD equations remains well-defined for all cases. 
\par
The first study of the MHD eigenvectors was carried out by Jeffrey and Tanuiti \citep{Jeffrey&Tanuiti1964} and was continued by others. The derived set of eigenvectors by Jeffrey and Tanuiti does not form a complete set near the points $B_x=0$ or ${B_y}^2+{B_z}^2=0$. Since in the proximity of the abovementioned conditions, this set of eigenvectors is not well defined as its columns become singular \citep{Brio&Wu1988}. To address this issue, a set of normalization parameters was proposed by Brio and Wu \citep{Brio&Wu1988}, and a new set of eigenvectors was derived for the cartesian coordinate system. These normalization parameters are $\alpha_{\mathrm{s}}$, $\alpha_{\mathrm{f}}$, and $\beta_{x(y,z)}$, defined in Eq. (\ref{eq:MHDeigen8}) \citep{Roe&Balsara1996}. These normalization parameters were proposed by others \citep{Roe&Balsara1996,Zachary&Colella1992,Dai&Woodward1994} as well. For these parameters, the following relations are held
\begin{equation}\label{eq:B1}
    {\alpha_{\mathrm{f}}}^2+{\alpha_{\mathrm{s}}}^2=1,\        \ {\alpha_{\mathrm{f}}}^2 {c_{\mathrm{f}}}^2+{\alpha_{\mathrm{s}}}^2 {c_{\mathrm{s}}}^2=a^2, \        \ {\alpha_{\mathrm{s}}}^2 {c_{\mathrm{f}}}^2 + {\alpha_{\mathrm{f}}}^2 {c_{\mathrm{s}}}^2 = b^2, \        \ \alpha_{\mathrm{f}} \alpha_{\mathrm{s}}=\frac{a \,b_{\perp}}{{c_{\mathrm{f}}}^2-{c_{\mathrm{s}}}^2},    
\end{equation}
where ${c_{\mathrm{f},\mathrm{s}}}^2=1/2\left[a^2+b^2 \pm \sqrt{\left(a^2+b^2 \right)^2-4 \, a^2 \,{b_x}^2}\right]$, $a^2=\gamma p/\rho$, $b={b_x}^2+{b_y}^2+{b_z}^2$, and ${b_{\perp}}^2={b_y}^2+{b_z}^2$.
Moreover, it can be shown that the square values of the fast and slow magnetosonic waves (${c_{\mathrm{f}}}^2$ and ${c_{\mathrm{s}}}^2$) are the bigger and smaller roots of the following equation \citep{Roe&Balsara1996}:
\begin{equation}\label{eq:B2}
    c^4 - (a^2+b^2)c^2+a^2{b_x}^2=0,
\end{equation}
where $c^2$ is an independent variable. Using Eq. (\ref{eq:B2}), we can easily prove that the following relations are satisfied:
\begin{equation}\label{eq:B3}
    {c_{\mathrm{f}}}^2 + {c_{\mathrm{s}}}^2 = a^2+b^2, \        \ c_{\mathrm{f}} \, c_{\mathrm{s}}=a \, |b_x|.
\end{equation}
 Eqs. (\ref{eq:B1}) and (\ref{eq:B3}) are widely used for simplifying the eigensystem of the MHD equations in the cartesian coordinate system \citep{Roe&Balsara1996}.\\
\indent In order to extend the MHD eigenstructure to the general three-dimensional curvilinear coordinate system, a new set of normalization parameters should be derived. In this study, these parameters are defined similar to the cartesian coordinates by substituting variables $a^2$, $b^2$, $b_{\perp}$, $c_{\mathrm{f}}$, $c_{\mathrm{s}}$, and $c_{\mathrm{a}}$ with the revised ones derived for the curvilinear case (see Eqs. (\ref{eq:MHDeigen4}) to  (\ref{eq:MHDeigen6})). Equations (\ref{eq:B1}) to (\ref{eq:B3}) are also held for the new defined normalization parameters in the general curvilinear coordinate system. In this work, these relations are widely used to simplify the set of calculated eigenvectors as much as possible and prevent any indeterminacy in the proposed set of eigensystem.


\bibliographystyle{plain}
\bibliography{main}

\end{document}